\documentclass[preprint,12pt]{elsarticle}

\usepackage{amssymb}

\usepackage{hyperref}
\usepackage{color}
\usepackage{subcaption}
\usepackage{graphicx}
\usepackage{placeins}

\newcommand{\Noon}{\textbf{n$\mathbf{_O^O}$n}}

\newcounter{bla}

\journal{Computer Physics Communications}

\begin{document}

\begin{frontmatter}



\title{A generator of forward neutrons for ultra-peripheral collisions: \Noon}

\author[a]{M. Broz\corref{author}}
\author[a]{J. G. Contreras}
\author[b]{J. D. Tapia Takaki}

\cortext[author] {Corresponding author.\\\textit{E-mail address:} Michal.Broz@cern.ch}
\address[a]{Faculty of Nuclear Sciences and Physical Engineering,
Czech Technical University in Prague, Czech Republic}
\address[b]{Department of Physics and Astronomy, The University of Kansas, Lawrence, KS, USA}

\begin{abstract}
The study of photon-induced reactions in  collisions of  heavy nuclei at RHIC and the LHC has become  an important direction of the research program of these facilities in recent years. In particular, the production of vector mesons in ultra-peripheral collisions (UPC) has been intensively studied.  Owing to the intense photon fluxes, the two nuclei participating in such processes  undergo electromagnetic dissociation producing neutrons at beam rapidities. Here, we introduce the \Noon\ (pronounced noon) Monte Carlo program, which generates events containing such neutrons. \Noon\ is a ROOT based program that can be interfaced  with existing generators of vector meson production in UPC or with theoretical calculations of such photonuclear processes. \Noon\ can also be easily integrated with the simulation programs of the experiments at RHIC and the LHC.
\end{abstract}

\end{frontmatter}


{\bf PROGRAM SUMMARY}

\begin{small}
\noindent
{\em Program Title:}    \Noon                                      \\
{\em Licensing provisions}: GNU GPLv3 \\    
{\em Programming language:}           C++                         \\
{\em External routines:} The generator is based on ROOT     \\
{\em Nature of problem:} The electromagnetic fields of nuclei at RHIC and the LHC can be described as a flux of quasi-real photons. These photons may interact with one of the nucleus in the opposite beam.
There are events of interest where  two independent interactions occur, one involving a hard scattering and one from the exchange of soft photons. As a result of the latter, the nucleus get excited and upon de-excitation it may emit neutrons, which are boosted to beam rapidities. The program computes the  probability of neutron emission based on existing measurements and some mild modelling; it then generates neutrons in a per-event basis.\\
{\em Solution method:} The break-up probabilities are computed using existing data and stored in ROOT objects (graphs and histograms). Photon energies from the accompanying hard process, e.g. vector meson production, are loaded into the program and a catalogue of specific break-up probabilities is constructed. The number of neutrons emitted in the event is generated and the neutrons are produced and boosted into the laboratory frame.  The output is a TTree with a TClonesArray of TParticles per event, which can be easily interfaced to the simulation programs of the RHIC and LHC collaborations.\\
{\em Restrictions:} At the moment only emission from Pb is available.\\
 {\em References:} \url{https://github.com/mbroz84/noon} and references in this article.\\ 
\end{small}

\section{Introduction}
\label{sec:Introduction}

The strong electromagnetic field of the  particles circulating at the Relativistic Heavy Ion Collider (RHIC) and at the Large Hadron Collider (LHC), can be understood as a flux of quasi-real photons as proposed by E. Fermi~\cite{Fermi:1924tc,Fermi:1925fq}. As the flux intensity depends on the square of the electric charge of the particle, heavy ions at these facilities produce an intense flux of photons. For this reason,  these accelerators can be consider as  photon-nucleus and photon-photon colliders. If the incoming particles approach each other at impact parameters larger than the sum of their radii, and because the strong force is short range,  hadronic interactions are  suppressed and only photon-induced processes are possible. Such reactions are called ultra-peripheral collisions (UPC). See~\cite{Krauss:1997vr,Baur:2001jj,Bertulani:2005ru,Baltz:2007kq} for an in-depth review of the subject and~\cite{Contreras:2015dqa} for an overview of recent results at the LHC.

One process that has attracted a lot of attention is the coherent production of a vector meson. This is because it is expected to be particularly sensitive to the target structure and its evolution within quantum chromodynamics (QCD)~\cite{Ryskin:1992ui}.  As examples, one can mention the measurements at RHIC of 
$\rho^0$~\cite{Agakishiev:2011me,Adler:2002sc,Abelev:2007nb,Adamczyk:2017vfu} and ${\rm J}/\psi$~\cite{Afanasiev:2009hy} vector mesons, or the measurements at the LHC of $\rho^0$~\cite{Adam:2015gsa,Sirunyan:2019nog}, ${\rm J}/\psi$~\cite{Abelev:2012ba,Abbas:2013oua,Khachatryan:2016qhq,Acharya:2018jua,Acharya:2019vlb}, $\psi^\prime$~\cite{Adam:2015sia} and $\Upsilon$~\cite{Sirunyan:2018sav} production in UPC.

The electromagnetic fields of the nuclei at RHIC and the LHC are so intense that the cross section for the exchange of a soft photon that excites at least one of the nuclei, a process known as electromagnetic dissociation (EMD), is very large. This cross section has been measured by ALICE at the LHC~\cite{ALICE:2012aa}. EMD is implemented in the RELDIS~\cite{Pshenichnov:2001qd,Pshenichnov:2011zz} which describes reasonably well the measured data.

Indeed, these electromagnetic fields are so strong that  the probability that  the  particles participating in the production of a vector meson in UPC undergo an additional and independent   exchange of photons is very large. Such an interaction may excite the nuclei producing  neutrons which are very slow in the rest frame of the emitting nucleus. See e.g.~\cite{Pshenichnov:2011zz} for a review. At RHIC and the LHC these neutrons are strongly boosted and appear at beam rapidities. 

Since the probability of producing such neutrons is large, online triggers aiming to select UPC events have been implemented based on neutrons detected by Zero Degree Calorimeters. 
Interestingly, UPC events with a coherently produced vector meson accompanied by neutrons at beam rapidities have also been proposed as a probe to study QCD~\cite{Baltz:2002pp,Guzey:2013jaa}. In UPC events either of the incoming nucleus can serve as the photon source. For vector mesons produced at rapidities different from zero there is a two-fold ambiguity on the photon energy.  To study the energy dependence of this process it is important to have a way to effectively determine the photon energy.  The measurement of vector mesons at a given rapidity and in terms of the number of neutrons at beam rapidities has been proposed as a tool to disentangle the two photon energies~\cite{Baltz:2002pp,Guzey:2013jaa}. Theoretical predictions for these processes for the current conditions at the LHC have been recently reported in~\cite{Guzey:2016piu}.

To perform  measurements of UPC events accompanied by neutrons, as discussed in~\cite{Guzey:2016piu},  a Monte Carlo (MC) generator that produces events with neutrons at beam rapidities is highly desirable. There are MC programs like STARlight~\cite{Klein:2016yzr} that generate the vector mesons and provide cross sections for events with forward neutrons, but unfortunately, STARlight does not provide final state neutrons in the event output.

In this article, we present the generator \Noon (pronounced noon) which, apart from generating forward neutrons, can easily be integrated to the simulation software of the RHIC and LHC  collaborations. \Noon\ can take as input either events produced by MC generators like STARlight or it can also use theoretical predictions of vector meson photoproduction. 

\section{Theoretical formalism}
\label{sec:Formalism}
As mentioned above, owing to the large photon flux from nuclei at RHIC and the LHC, in UPC  there is a large probability  that photon exchanges will excite one or both of the colliding nuclei.
Assuming that the sub-processes are independent, the cross section of a UPC accompanied by the break-up of one or both nuclei is~\cite{Baltz:2002pp}:
\begin{equation}
\sigma(AA \rightarrow PA^\prime_i A^\prime_j) \propto \int d^{2}\vec{b} P_{P}(b) P_{ij}(b) \exp(-P_{H}(b)), 
\label{eq:1}  
\end{equation}
where $P$ denotes the final state produced by the hard process (in this study we assume a coherently produced vector meson), $\vec{b}$ denotes the impact parameter with $b$ its magnitude,  $i,j = 0,1,2\dots$ are the number of neutrons emitted by the nucleus on either side, and $A^\prime_{i,j}$ represents the ion after the neutron emission.   

 There are 3 probabilities in this equation:
\begin{itemize}
    \item[$(a)$] The probability of the hard photoproduction process, $P_{P}(b)$. 
    \item[$(b)$] The probability of nuclear break-up with emission of $i$ and $j$ neutrons from the first and second nucleus, respectively, $P_{ij}(b)$.
    \item[$(c)$] The probability of a hadronic interaction, $P_{H}(b)$. 
\end{itemize}

\subsection{The probability of no hadronic interaction}
\begin{figure}[t!]
\begin{center}
\begin{subfigure}[b]{0.49\textwidth}
         \includegraphics[width=\textwidth]{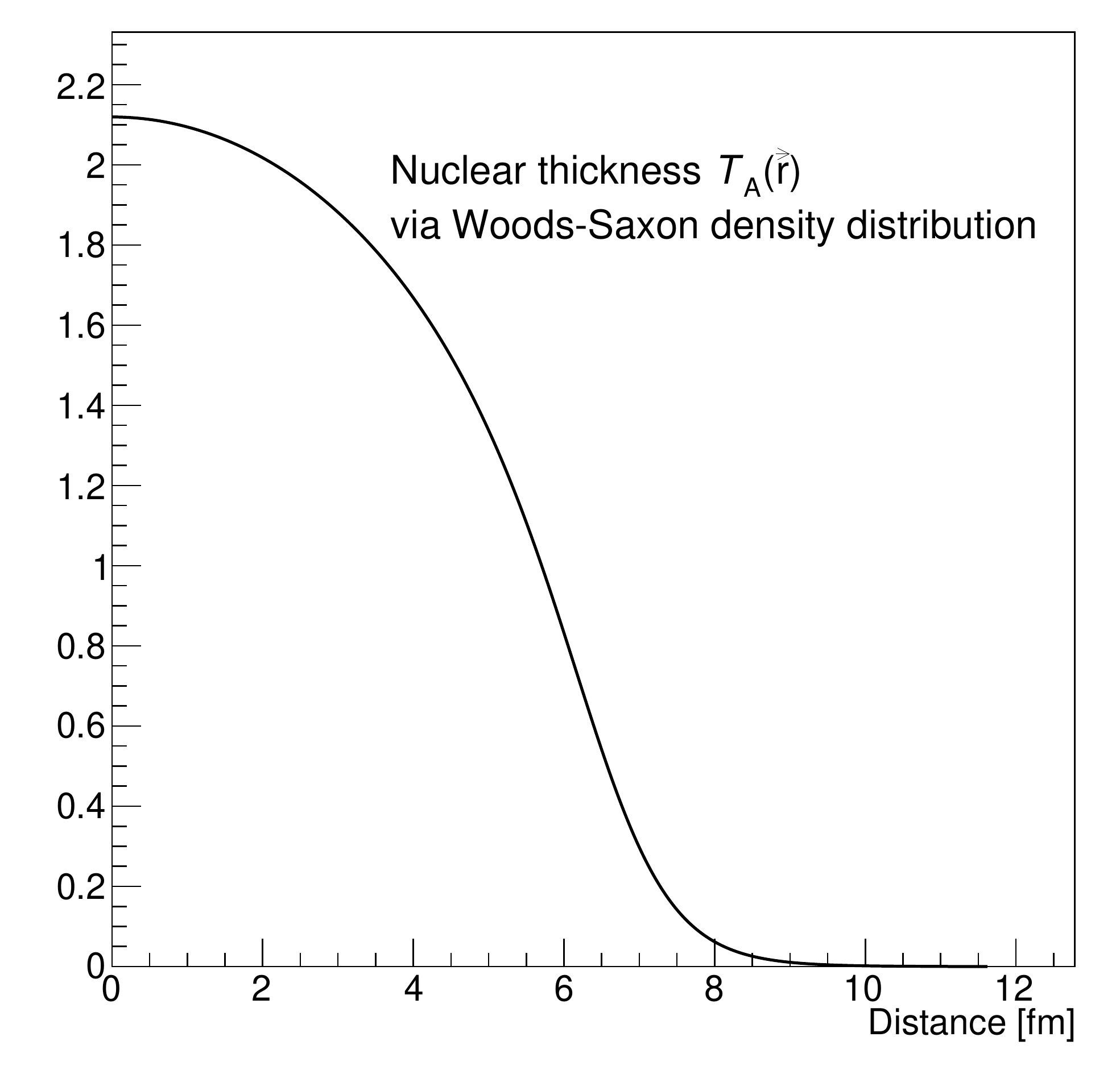}
         \caption{ }
         \label{fig:HadronicProb_a}
     \end{subfigure}
\begin{subfigure}[b]{0.49\textwidth}
         \includegraphics[width=\textwidth]{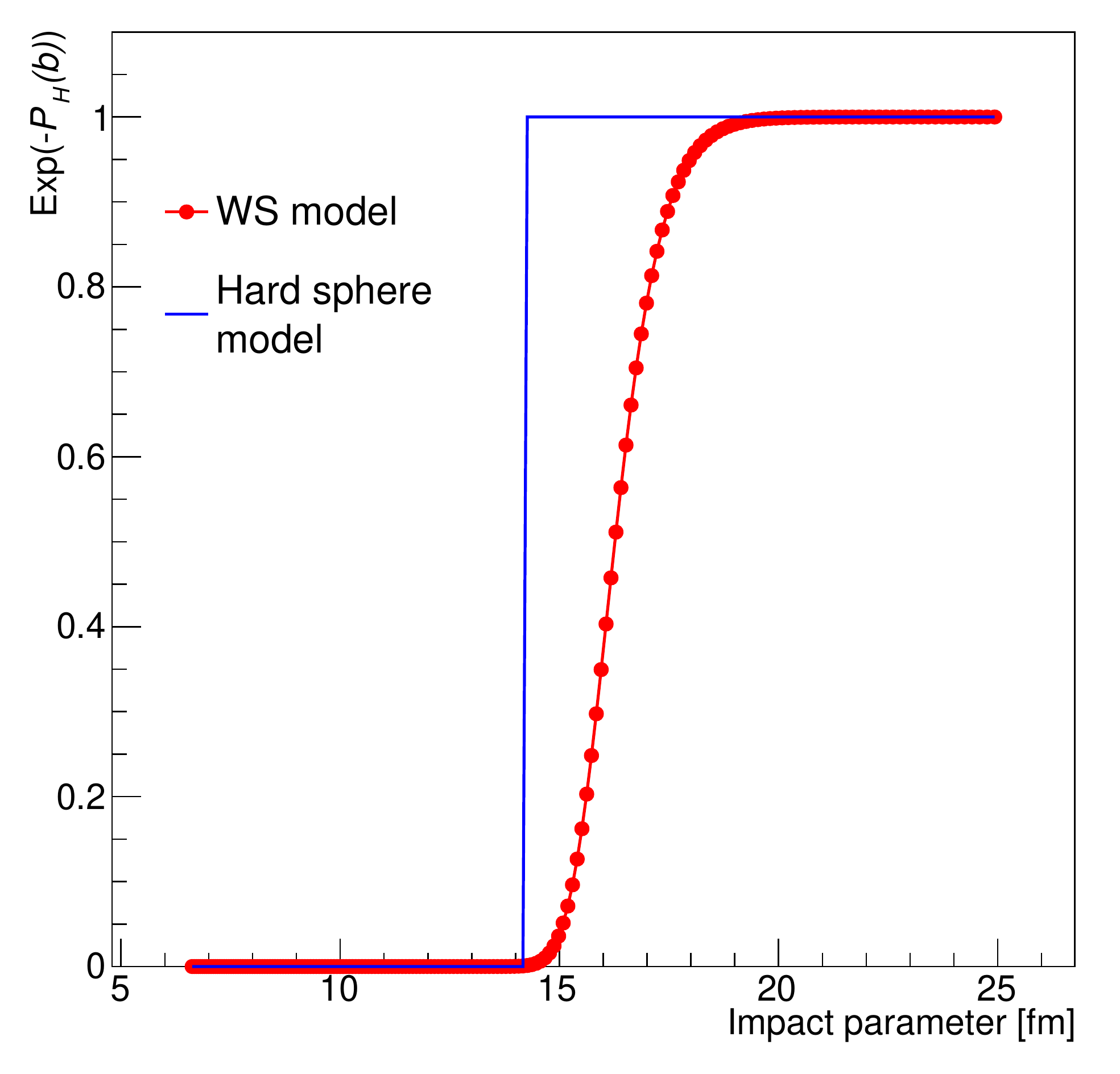}
         \caption{ }
         \label{fig:HadronicProb_b}
     \end{subfigure}
\caption{(Colour online) Nuclear density as a function of the distance from the centre of the $\rm{^{208}Pb}$ nucleus (a). Probability of no hadronic interaction between two nuclei as a function of the impact parameter (b). The red line corresponds to a WS distribution, while in the hard-sphere approximation $\rho_{A}(s)$ is given by a Heaviside function.}
\label{fig:HadronicProb}
\end{center}
\end{figure}

The factor $\exp(-P_{H}(b))$ in Eq.~(\ref{eq:1}) ensures that the reaction is unaccompanied by hadronic interactions. Therefore, in this work we only consider the Coulomb break-up of the nucleus. $P_{H}(b)$, the mean number of  nucleons in the $A$ and $B$ incoming nuclei that interact at least once, is given by
\begin{equation}
P_{H}(b) = \int d^{2}\vec{r} T_{\rm{A}}(\vec{r} -\vec{b}) (1-\exp(-\sigma_{\rm{NN}}T_{\rm{B}}(\vec{r}))), 
\label{eq:2}  
\end{equation}
where $\sigma_{\rm{NN}}$ is the nucleon-nucleon inelastic cross section at the corresponding centre-of-mass energy, $T_{\rm{A}}(\vec{r})$ is the nuclear thickness function
\begin{equation}
T_{\rm{A}}(\vec{r}) = \int dz \rho_{A}(\sqrt{|\vec{r}|^{2} + z^{2}}), 
\label{eq:3}  
\end{equation}
and $\rho_{A}(s)$ is the nuclear density for the nucleus $A$ at a distance $s$ from its centre; it can be modelled, for example, with a Woods-Saxon (WS) distribution.  For symmetric nuclei 
\begin{equation}
\rho_{A}(s) = \frac{\rho_0}{1+\exp(\frac{s-R_{\rm{WS}}}{d})},
\label{eq:4}  
\end{equation}
with $R_{\rm{WS}}$ and $d$ appropriate parameters. Figure~\ref{fig:HadronicProb} shows the impact-parameter dependence of this probability and of the nuclear thickness for $\rm{^{208}Pb}$, which is used at the LHC.

\subsection{Probability of the hard process}
The probability  $P_{P}(b)$ is given by
\begin{equation}
P_{P}(b) = \int dk \frac{d^{3}n(b,k)}{dkd^2\vec{b}}\sigma_{\gamma A \rightarrow PA}(k), 
\label{eq:5}  
\end{equation}
where $\sigma_{\gamma A \rightarrow PA}(k)$ is the cross section of the photonuclear process $P$, and the flux of photons produced with energy $k$ at $b$, in the semi-classical approximation~\cite{Krauss:1997vr,Baur:2001jj,Bertulani:2005ru,Baltz:2007kq}, can be expressed by:
\begin{equation}
\frac{d^{3}n(b,k)}{dkd^{2}\vec{b}} = \frac{\rm{Z^2} \alpha}{\pi^{2} \gamma^{2}} k \Bigg[K^{2}_{1}(\frac{kb}{\gamma}) + \frac{1}{\gamma^{2}}K^{2}_{0}(\frac{kb}{\gamma})\Bigg],
\label{eq:6}  
\end{equation}
where $\gamma$ is the Lorentz factor, $Z$ the electric charge of the nucleus, $\alpha$ the electromagnetic coupling constant and $K_1$ a Bessel function. The combination of Eq.~(\ref{eq:1}) and Eq.~(\ref{eq:5}) yields:
\begin{equation}
\sigma(AA \rightarrow PA^\prime_i A^\prime_j)  \propto \int d^{2}\vec{b} \int dk \frac{d^{3}n(b,k)}{dkd^{2}\vec{b}} \sigma_{\gamma A \rightarrow PA}(k) P_{ij}(b) \exp(-P_{H}(b)). 
\label{eq:7}  
\end{equation}

In a given interaction the value of $k$ is fixed and for exclusive interactions it can be derived from the kinematics of the final state. For the case of the coherent production of a vector meson  $k$ depends only on the invariant mass ($M$) and rapidity ($y$) of the produced vector meson:
\begin{equation}
k = \frac{1}{2}M_{V}\exp(\pm y).
\label{eq:8}  
\end{equation}
For that value of $k$, $\sigma_{\gamma A \rightarrow PA}(k)$ can be treated as a constant and  Eq.~(\ref{eq:1}) can be rewritten in  a simpler form:
\begin{equation}
\left. \sigma(AA \rightarrow PA^\prime_i A^\prime_j) \right|_{k =\rm{const}} \propto \int d^{2}\vec{b} \frac{d^{3}n(b,k)}{dkd^{2}\vec{b}} P_{ij}(b) \exp(-P_{H}(b)). 
\label{eq:9}  
\end{equation}
Then, the probability that a photoproduction event $P$, associated with a photon of energy $k$, is accompanied by nuclear break-up can be defined as:
\begin{equation}
\left. P(AA \rightarrow A^\prime_i A^\prime_j) \right|_{k =\rm{const}}= \frac{\int d^{2}\vec{b} \frac{d^{3}n(b,k)}{dkd^{2}\vec{b}} \exp(-P_{H}(b)) P_{ij}(b)}{\int d^{2}\vec{b} \frac{d^{3}n(b,k)}{dkd^{2}\vec{b}} \exp(-P_{H}(b))}. 
\label{eq:10}  
\end{equation}

\subsection{Nuclear break-up probability}
The probability for nuclear break-up, under the assumption of independent nuclear break-up, can be factorised as the product of the break-up probabilities of each nucleus:
\begin{equation}
P_{ij}(b) = P_{i}(b) \times P_{j}(b).
\label{eq:11}  
\end{equation}

Let  $P_{\rm{Xn}}$ be  the probability of  nuclear break-up of one nucleus to a state with any number (X) of neutrons (n).  Under the assumption of a Poisson distribution, the probability of having exactly $L$ neutrons is:
\begin{equation}
P_{\rm{Ln}}(b) = \frac{(P^{1}_{\rm{Xn}}(b))^{L} \times \exp(-P^{1}_{\rm{Xn}}(b))}{L!},
\label{eq:12}  
\end{equation}
while the probability to have at least one excitation is:
\begin{equation}
P_{\rm{Xn}}(b) = 1-\exp(-P^{1}_{\rm{Xn}}(b)).
\label{eq:13}  
\end{equation}
Here $P^{1}_{\rm{Xn}}(b)$ is the mean number of the Coulomb excitations of the nucleus to any state which emits one or more neutrons. It has a similar form to that of $P_{P}(b)$, shown in Eq.~(\ref{eq:5}): 
\begin{equation}
P^{1}_{\rm{Xn}}(b) = \int dk \frac{d^{3}n(b,k)}{dkd^{2}b}\sigma_{\gamma A \rightarrow A'+ \rm{Xn}}(k).
\label{eq:14}  
\end{equation}
\begin{figure}[t!]
\begin{center}
\includegraphics[width=\textwidth]{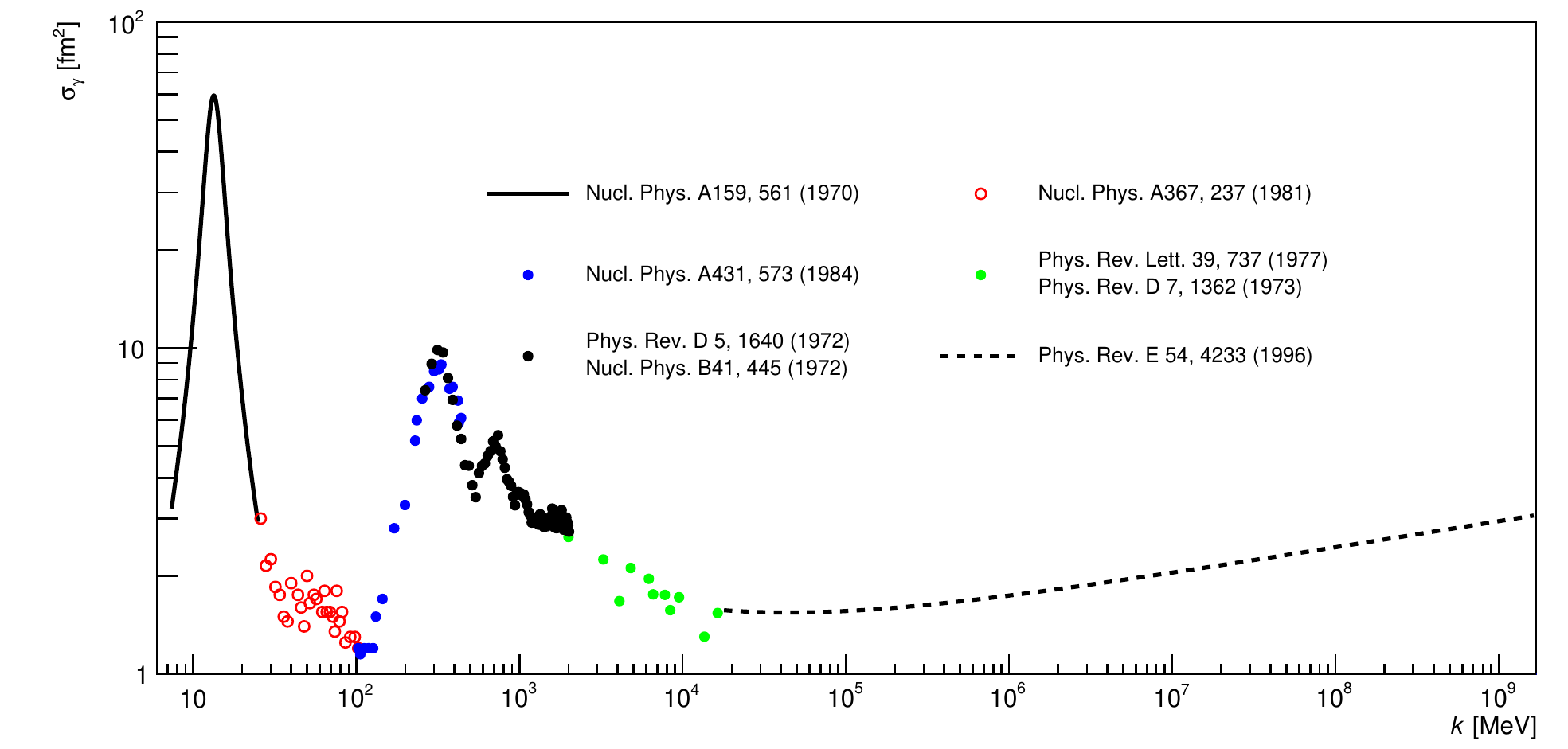}
\caption{(Colour online) The cross section $\sigma_{\gamma A \rightarrow A'+ \rm{Xn}}(k)$ for $\rm{^{208}Pb}$. Various experiments and approaches are used to describe different energy ranges. See text for details.} 
\label{fig:xSectionXn}
\end{center}
\end{figure}

\begin{figure}[t!]
\begin{center}
\includegraphics[width=\textwidth]{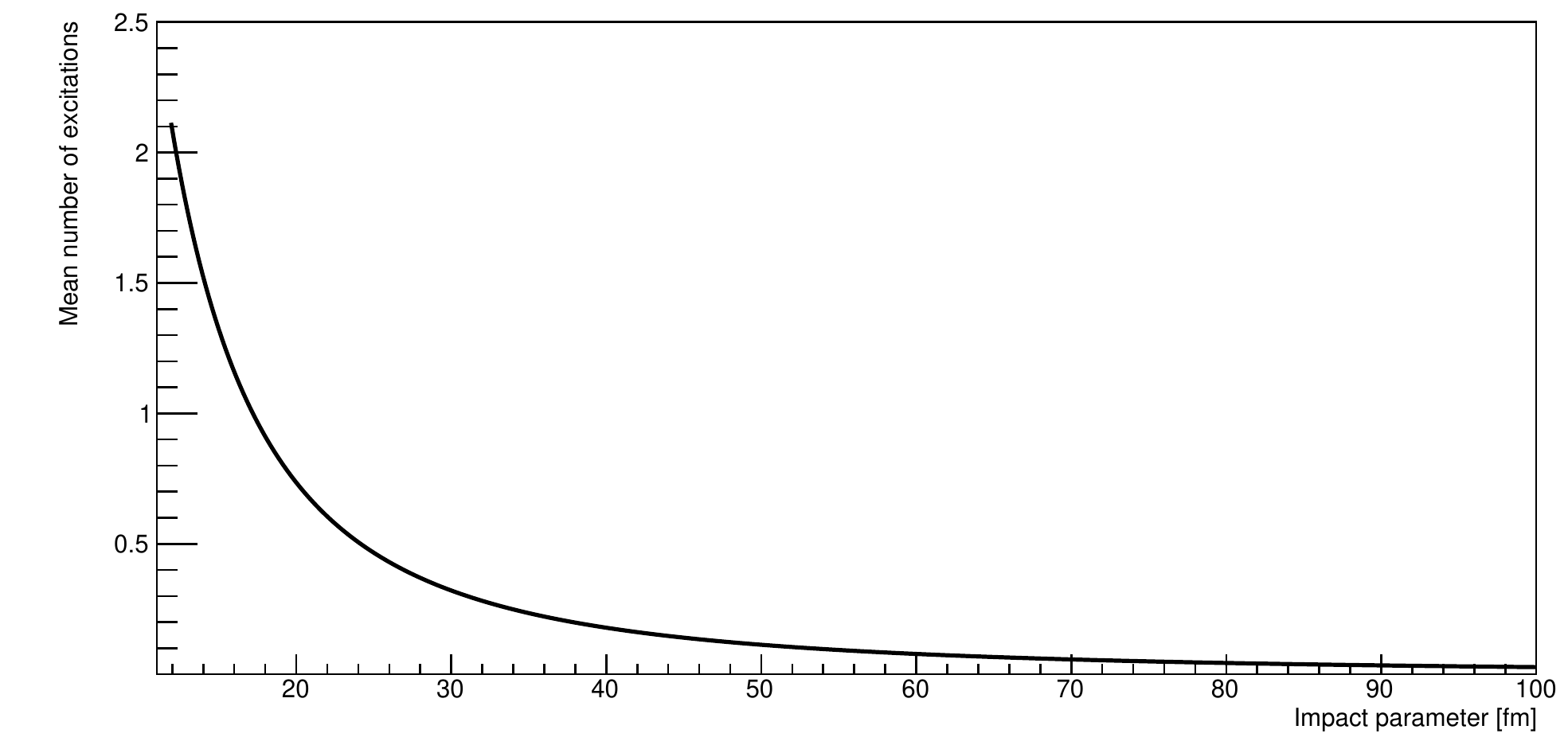}
\caption{ $P^{1}_{\rm{Xn}}(b)$ from Eq.~(\ref{eq:14}) for the case of $^{208}$Pb.}
\label{fig:ProbXn}
\end{center}
\end{figure}

The cross section $\sigma_{\gamma A \rightarrow A'+ \rm{Xn}}(k)$ is shown in Fig.~\ref{fig:xSectionXn}. It is determined using experimental data from fixed-target experiments covering the range of photon energies up to $k<16.4$ GeV~\cite{Veyssiere:1970ztg,Lepretre:1981tf,Carlos:1985te,Armstrong:1971ns,Armstrong:1972sa,Michalowski:1977eg,Caldwell:1973bu,Caldwell:1978ik}.  The data of~\cite{Veyssiere:1970ztg} have been corrected according to~\cite{Berman:1987zz}. For larger energies the Regge theory parametrisation from~\cite{Baltz:1996as,Baltz:1998ex} is used. The impact-parameter dependence of $P^{1}_{\rm{Xn}}(b)$ computed using this cross section is shown in Fig.~\ref{fig:ProbXn}.

\begin{figure}[t!]
\begin{center}
\includegraphics[width=\textwidth]{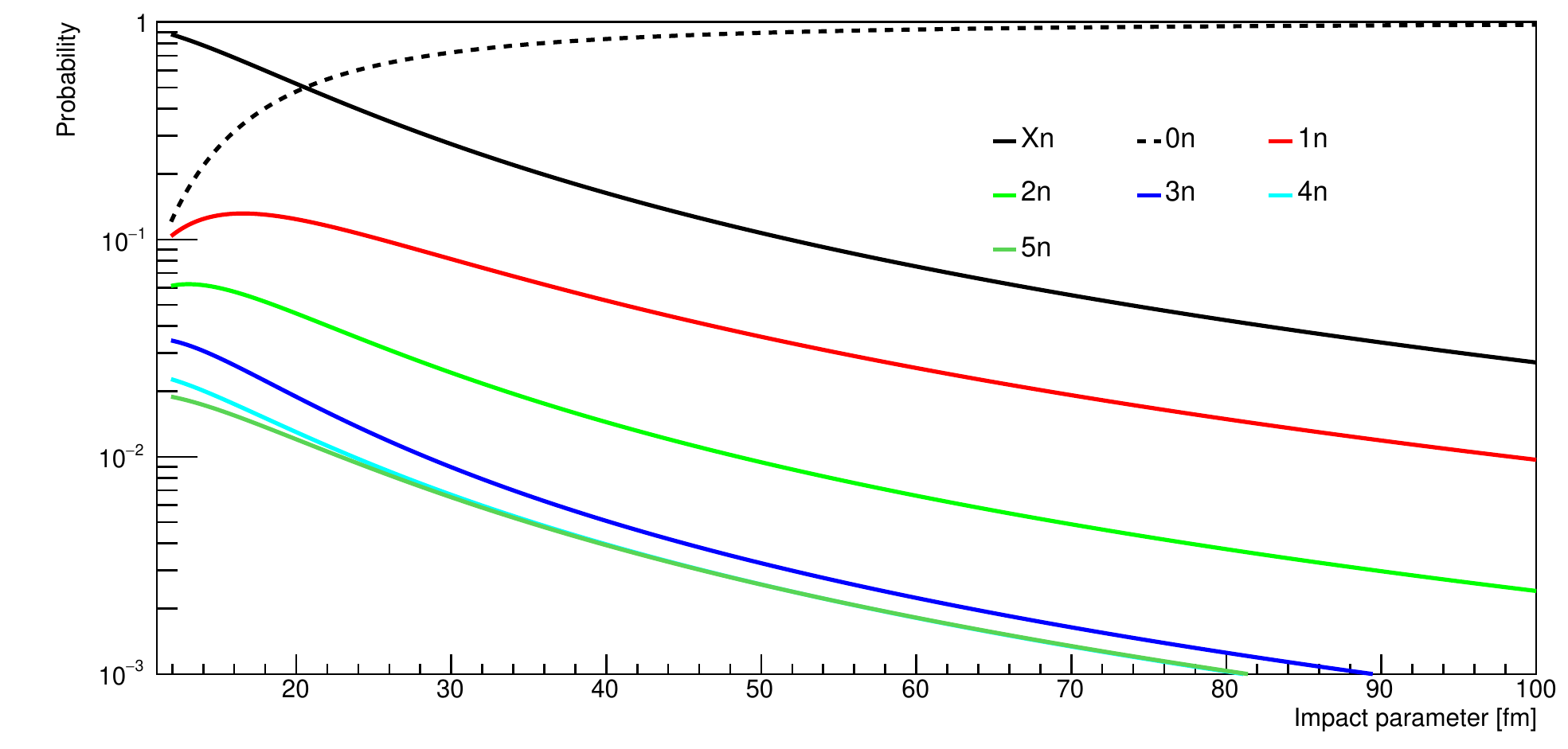}
\caption{(Colour online) Nucleus break-up probabilities for  $^{208}$Pb as a function of impact parameter $P_{i}(b)$ for various neutron multiplicities including the no-break-up scenario and break-up to any number of neutrons.}
\label{fig:Breakup}
\end{center}
\end{figure}

In a similar way the probability of a nucleus going into a state with N neutrons is:
\begin{equation}
P^{1}_{\rm{Nn}}(b) = \int dk \frac{d^{3}n(b,k)}{dkd^{2}b}\sigma_{\gamma A \rightarrow A'+ \rm{Nn}}(k),
\label{eq:15}  
\end{equation}
such that
\begin{equation}
P^{1}_{\rm{Xn}}(b) = \sum_{N=1}^{\infty}P^{1}_{\rm{Nn}}(b).
\label{eq:16}  
\end{equation}
Explicitly, the first few terms are 
\begin{eqnarray}
P_{1}(b) & = & P^{1}_{\rm{1n}}(b) \times \exp(-P^{1}_{\rm{Xn}}(b)), \\
P_{2}(b) & = & [P^{1}_{\rm{2n}}(b) + \frac{(P^{1}_{\rm{1n}}(b))^2}{2!}] \times \exp(-P^{1}_{\rm{Xn}}(b)), \\
P_{3}(b) & = & [P^{1}_{\rm{3n}}(b) + 2P^{1}_{\rm{2n}}(b)P^{1}_{\rm{1n}}(b) + \frac{(P^{1}_{\rm{1n}}(b))^3}{3!}] \times \exp(-P^{1}_{\rm{Xn}}(b)). 
\label{eq:17}  
\end{eqnarray}
This means for example, that two neutron states can be produced either by direct two neutron emission or by two emissions of one neutron; contributions to the three neutron states are from three one-neutron emissions, one emission of three neutrons, or emissions of one and two neutrons.
The sum of these higher order excitations and their various combinations form together $P_{\rm{Xn}}(b)$ from Eq.~(\ref{eq:13}). Fortunately, since  $P^{1}_{\rm{Xn}}(b)$ is small  even at small impact parameters, to integrate more than 99$\%$ of the probability, it is enough to sum up to 6 excitations and the probability of higher-order contributions falls down very quickly, as shown in  Fig.~\ref{fig:Breakup}.

In principle, the unitarity condition
\begin{equation}
\sum_{N=1}^{\infty}P_{\rm{Nn}}(b) = P_{\rm{Xn}}(b)
\label{eq:18}  
\end{equation}
should be fulfilled. At the same time, its implementation is not trivial. Higher order excitations can easily produce neutron multiplicities which are out of the region of interest,  or are  non-physical since they exceed the number of neutrons in the nucleus. By default the generator computes break-up probabilities up to 50 neutrons and 5 excitations, which is enough for the applications to RHIC and LHC that we have in mind.

\begin{figure}[t!]
\begin{center}
\includegraphics[width=\textwidth]{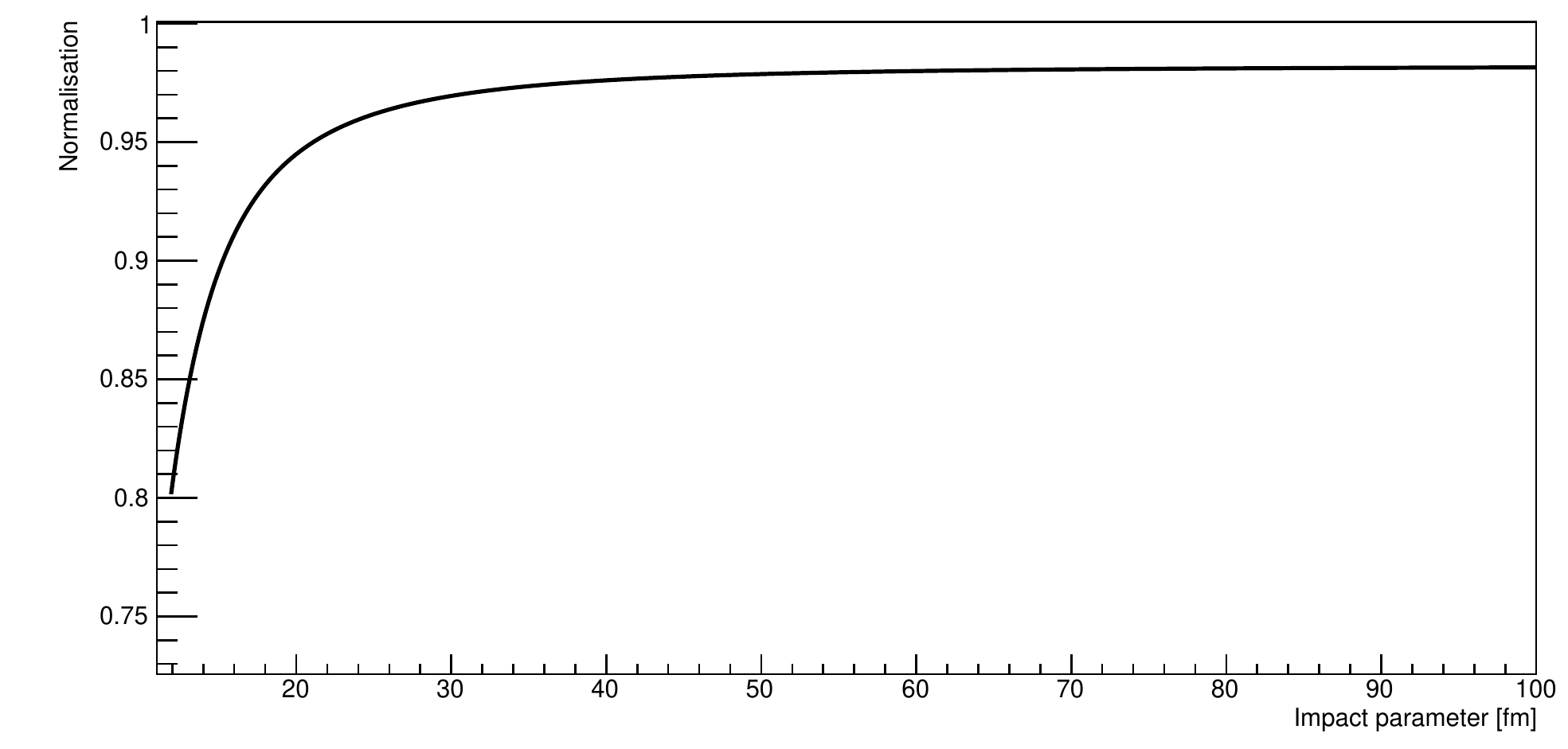}
\caption{Normalisation ratio $(\sum_{N=1}^{\infty}P_{\rm{Nn}})/P_{\rm{Xn}}$ as a function of impact parameter $b$. See text for details.}
\label{fig:Unity}
\end{center}
\end{figure}

The sum $\sum_{N=1}^{\infty}P_{\rm{Nn}}(b)$ is renormalised at every $b$ step to be equal to $ P_{\rm{Xn}}(b)$. The renormalisation constant, given by the ratio $\frac{\sum_{N=1}^{\infty}P_{\rm{Nn}}}{P_{\rm{Xn}}}$, is shown in Fig.~\ref{fig:Unity}. It is around 0.8 at small impact parameters, while at 30 fm it reaches a plateau at around 0.98. It is worth to mention that increasing the number of excitations to 6 or more would not significantly change the predictions of \Noon.       

\section{Generation of neutron multiplicity and energy}
\label{sec:NeutronMultiplicity}
\subsection{Generation of neutron multiplicity}
To compute the Coulomb break-up of the nucleus to a defined number of neutrons using Eq.~(\ref{eq:16}) and Eq.~(\ref{eq:17}) the partial cross sections $\sigma_{\gamma A \rightarrow A'+ \rm{Nn}}(k)$ are needed for the full photon energy range covered at the LHC.
Note that the cross sections  naturally decrease very fast with increasing number of neutrons. Furthermore, taking into account the precision of the current and expected UPC measurements it is justified to  make some reasonable  approximations.    

\begin{figure}[t!]
\begin{center}
\includegraphics[width=\textwidth]{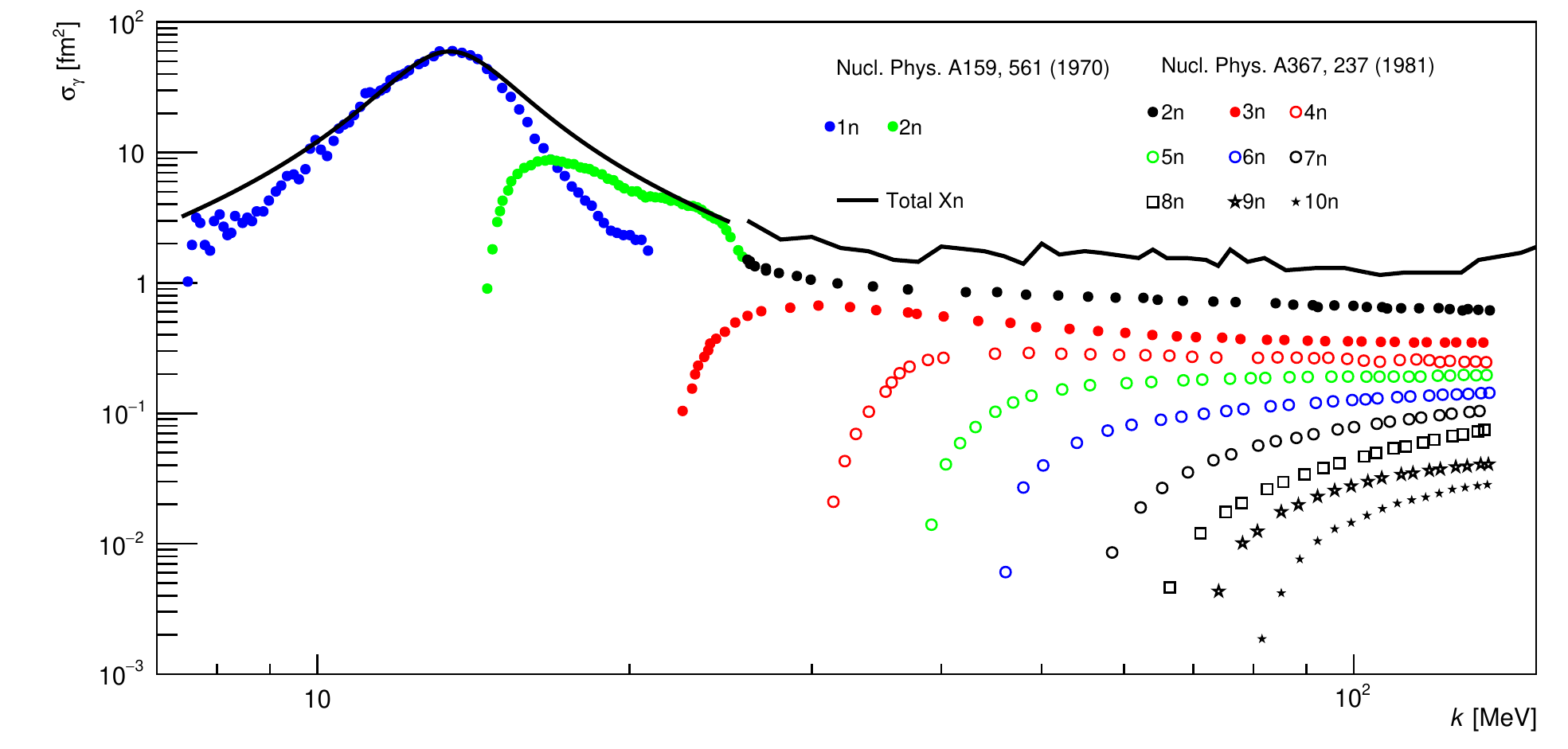}
\caption{(Colour online) Partial cross section for $\rm{^{208}Pb}$ and various neutron multiplicities as measured by the experiments quoted on the figure.}
\label{fig:xSectionNn}
\end{center}
\end{figure}

Clearly, the most important region of the total cross section is the Giant Dipole Resonance (GDR) peak at low energies. The GDR excitations produce mainly final states with one or two neutrons and were investigated in detail by various experiments. The measurement of partial cross sections, up to 10 neutrons and up to 140 MeV, is reported in~\cite{Lepretre:1981tf}. Thus, in the low-energy region these precise data, which are  shown in Fig.~\ref{fig:xSectionNn}, are used.

The same collaboration used the partial cross sections to extract the average and the dispersion of the number of neutrons as a function of the incident photon energy~\cite{Lepretre:1982xs}. Various approaches based on different motivations and background considerations were used to extrapolate these averages to higher energies~\cite{ArrudaNeto:1977uf}. One of them, was followed here: the average and dispersion, as a function of photon energy, was fitted to a logarithm and extrapolated to higher energies. Even though the range of the extrapolation here is rather large---the fit covers one order of magnitude in energy, up to 140 MeV, and the extrapolation is up to $10^9$ MeV---the fact that the evolution is logarithmic helps to stabilise the numerical results.

\begin{figure}[t!]
\begin{center}
\includegraphics[width=\textwidth]{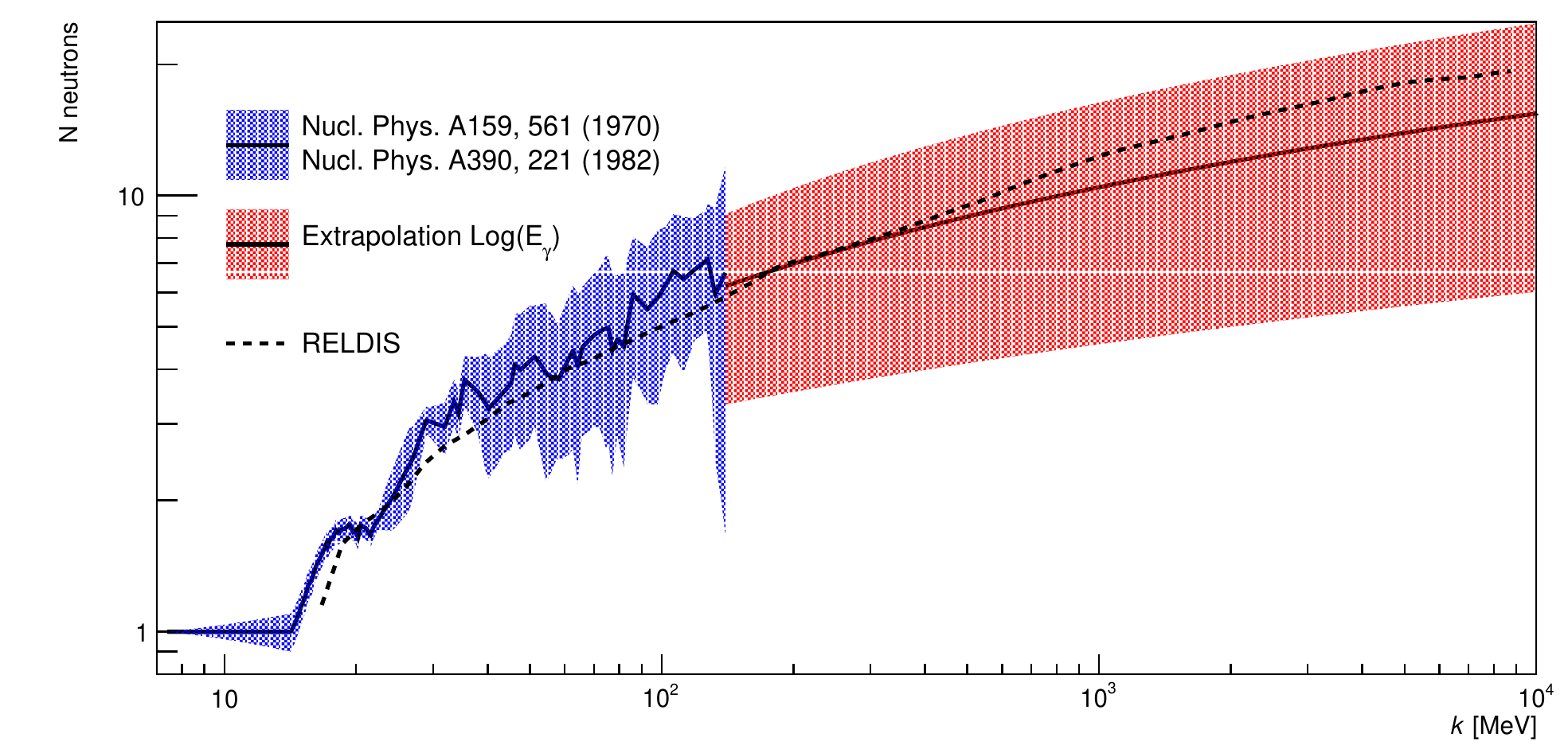}
\caption{(Colour online) Arithmetic average (line) and dispersion (dashed area) of neutron multiplicity as a function of the incident photon energy. The approach used here and the prediction of the RELDIS model~\cite{Pshenichnov:2011zz} are  shown in red and with a dash line, respectively.}
\label{fig:MultiExtrapol}
\end{center}
\end{figure}

We found that this approach seems to work rather well in describing the neutron multiplicity as a function of the photon energy. Figure~\ref{fig:MultiExtrapol} shows both the data used for the fit and the logarithmic extrapolation function. The line shows the average and the band represents the dispersion. This approach is compared with results of the RELDIS model~\cite{Pshenichnov:2011zz} and found in rather good agreement.        

\begin{figure}[t!]
\begin{center}
\begin{subfigure}[b]{0.49\textwidth}
         \includegraphics[width=\textwidth]{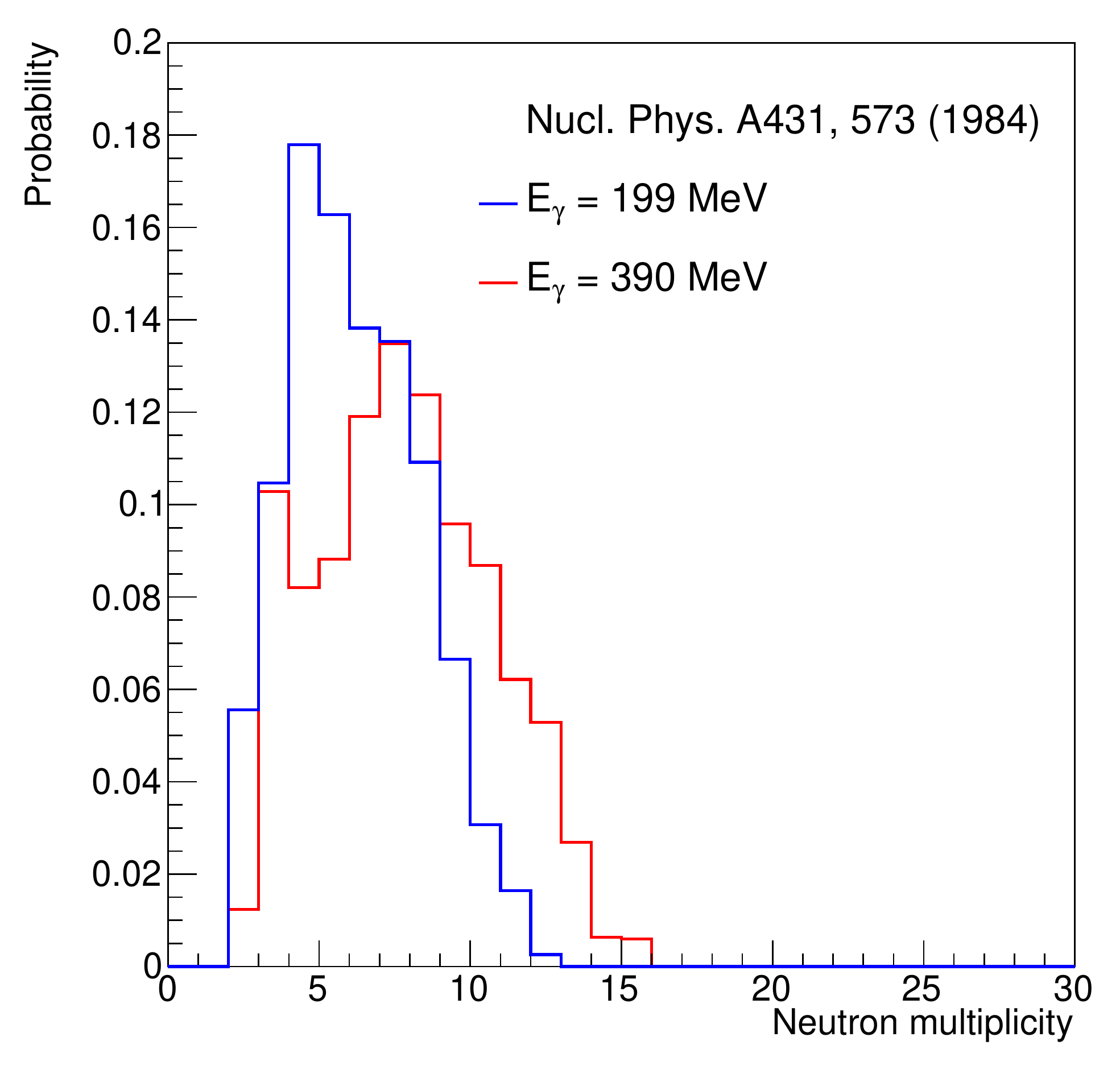}
         \caption{ }
         \label{fig:MultiUnfold_a}
     \end{subfigure}
\begin{subfigure}[b]{0.49\textwidth}
         \includegraphics[width=\textwidth]{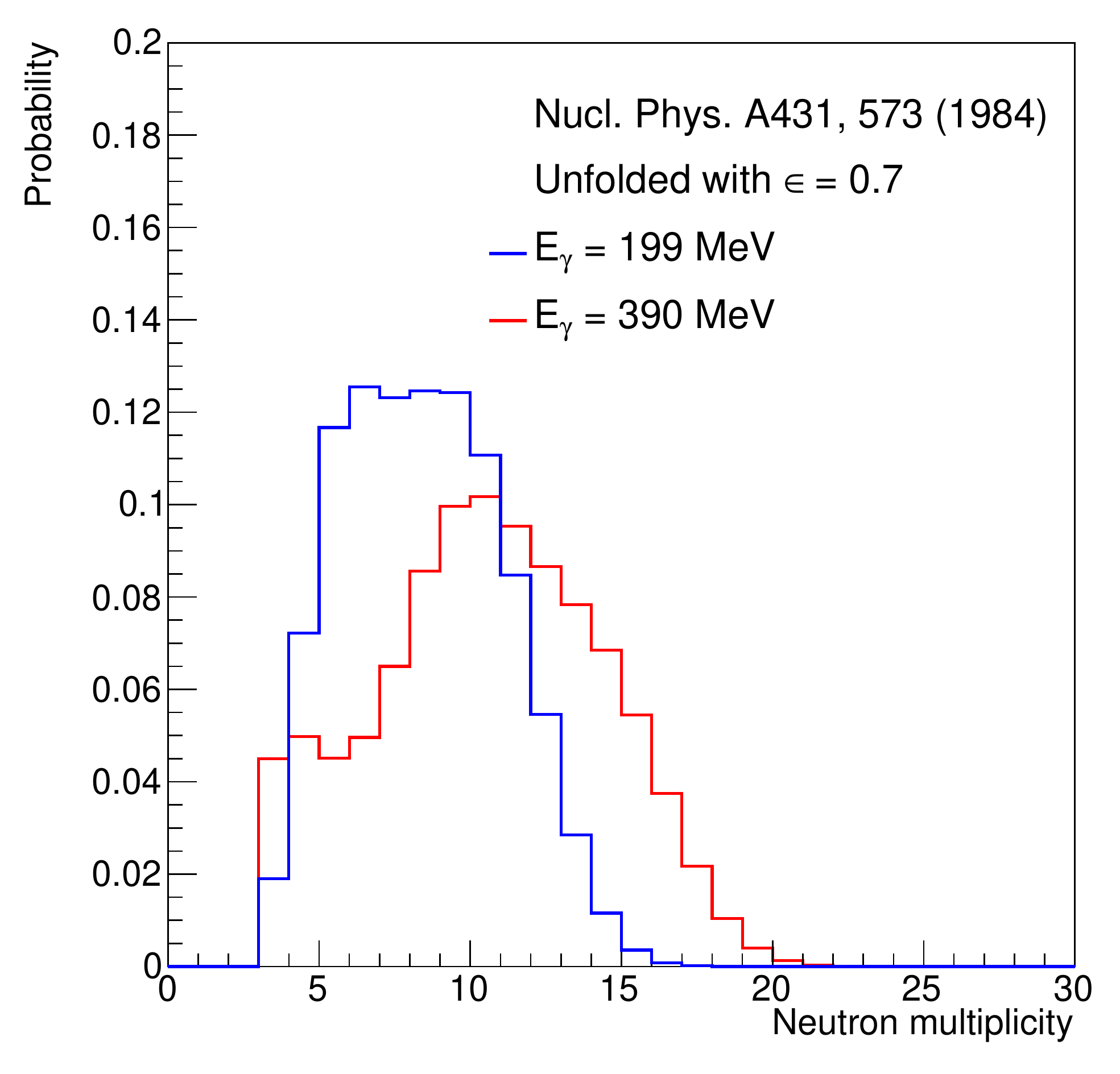}
         \caption{ }
         \label{fig:MultiUnfold_b}
     \end{subfigure}
\caption{(Colour online) Multiplicity distribution as measured by~\cite{Carlos:1985te} at photon energies of 199 MeV and 390 MeV (a). Unfolded multiplicity distributions (b).}
\label{fig:MultiUnfold}
\end{center}
\end{figure}

As  mentioned above, no measurement of the partial cross sections or neutron multiplicities exists above 140 MeV. Nonetheless, information on the shape of the neutron multiplicity distribution at photon energies of 199 MeV and 390 MeV is reported in~\cite{Carlos:1985te}. Figure~\ref{fig:MultiUnfold_a} shows the shape of the neutron-multiplicity distribution as measured by~\cite{Carlos:1985te}. These shapes are convoluted with the efficiency of the neutron detectors used in the experiment ($\epsilon = 0.7$). Using a binomial response matrix, with the singe hit efficiency mentioned in~\cite{Carlos:1985te}, the shapes have been de-convoluted and found to be in a rather good agreement with Gaussian shapes as shown in Fig.~\ref{fig:MultiUnfold_b} .    

\begin{figure}[t!]
\begin{center}
\includegraphics[width=\textwidth]{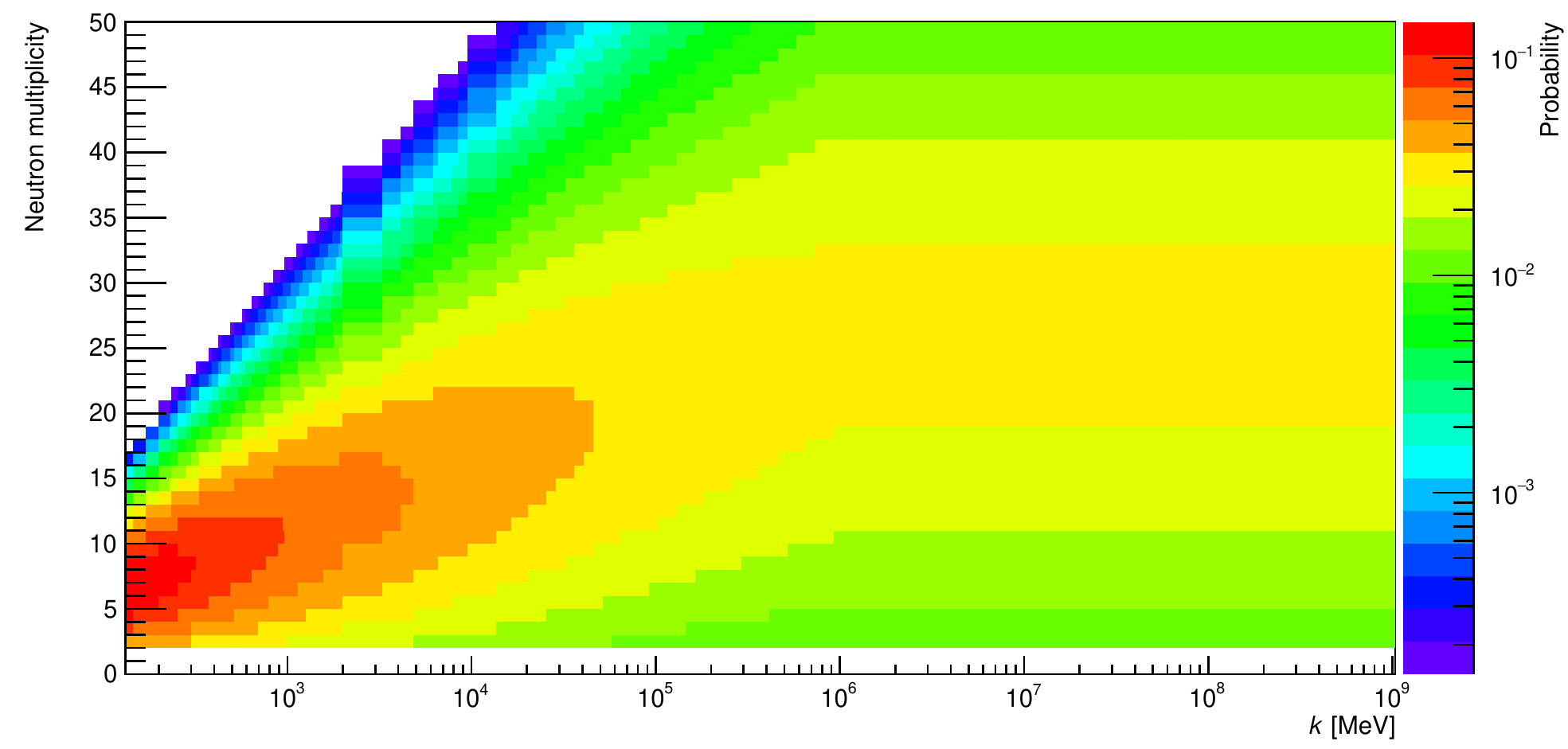}
\caption{(Colour online) Branching ratio (colour scale) of the total cross section to the various neutron multiplicities. Branching ratios are normalised to unity for every energy bin. Binning in energy is non-uniform and corresponds to binning of the total  cross section shown in Fig.~\ref{fig:xSectionXn}.}
\label{fig:MultiBRs}
\end{center}
\end{figure}

In summary, the branching ratios to each partial cross section are computed from the fit by extrapolating the arithmetic average and dispersion  shown in Fig.~\ref{fig:MultiExtrapol}, and using a Gaussian approximation for the shape. A map of the branching ratios as a function of the incident photon energy is shown in Fig.~\ref{fig:MultiBRs}. 

\subsection{Neutron energy generation}
Measurement of the spectra of secondary particles from a mono-energetic source of photons are currently limited. Fortunately, the neutron energy dependence strongly influences the relative population of various product nuclides when multi-particle production is possible.  Thus, one can rely on the accuracy of the evaluated spectra when the measured and calculated cross sections are in good agreement. 

We used the emission spectra of the secondary particles from the Photonuclear Data for Applications project of the International Atomic Energy Agency~\cite{Chadwick:2011xwu}. Tables are available in the Evaluated Nuclear Data File (ENDF) format~\cite{osti_981813}.

\begin{figure}[t!]
\begin{center}
\begin{subfigure}[b]{0.49\textwidth}
         \includegraphics[width=\textwidth]{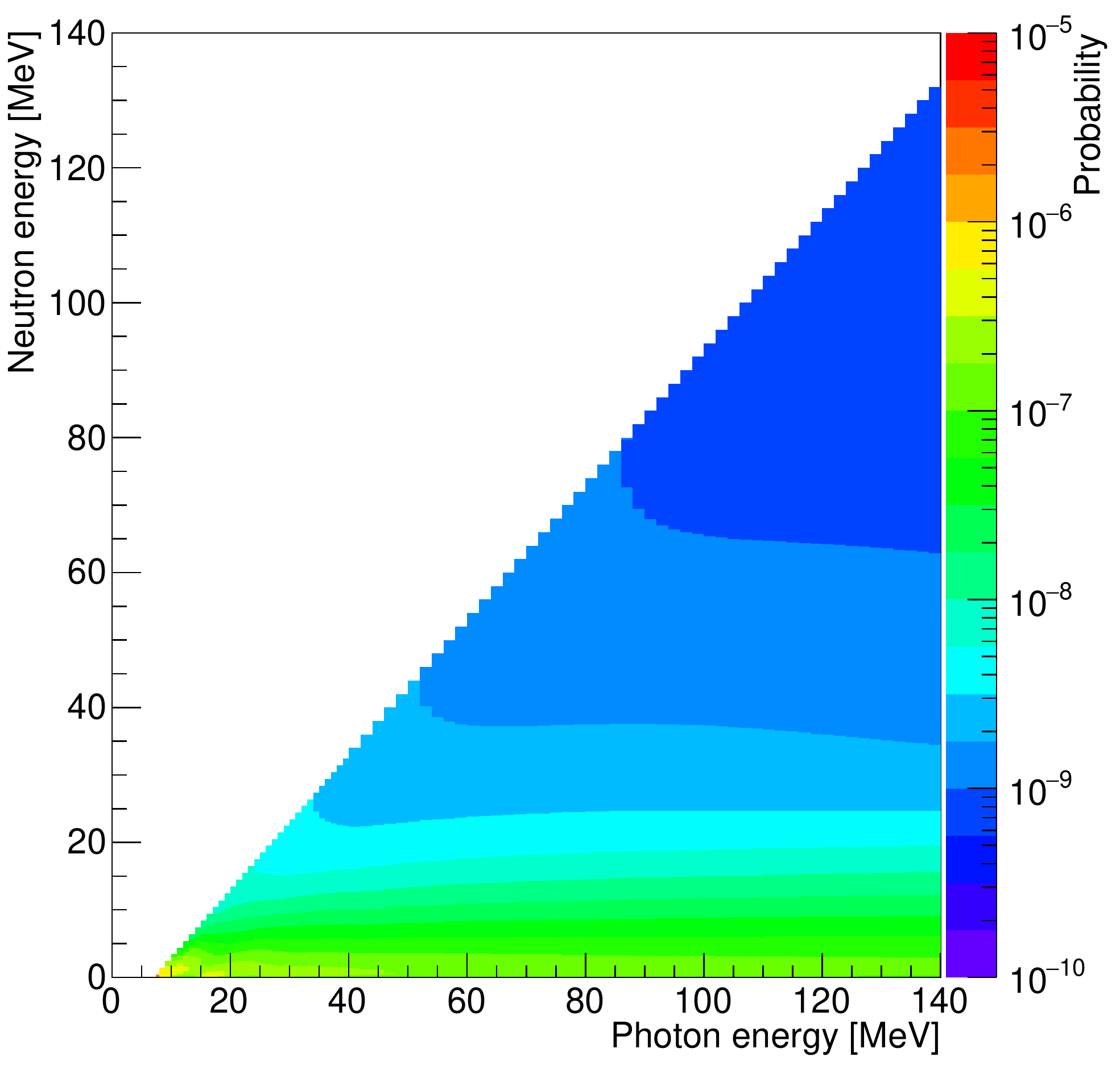}
         \caption{ }
         \label{fig:ENDF_a}
     \end{subfigure}
\begin{subfigure}[b]{0.49\textwidth}
         \includegraphics[width=\textwidth]{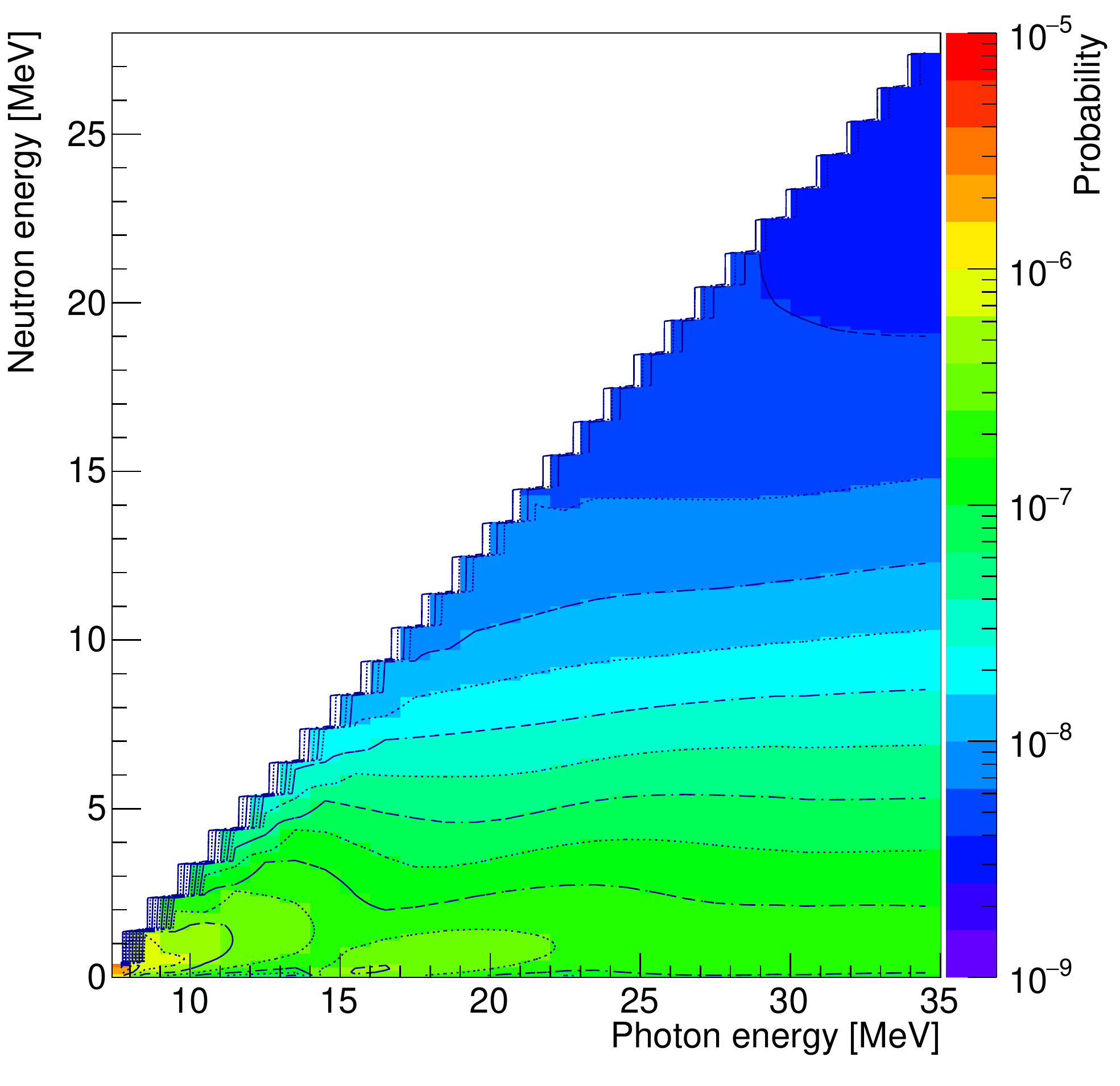}
         \caption{ }
         \label{fig:ENDF_b}
     \end{subfigure}
\caption{(Colour online) Evaluated nuclear data. Emission spectra of secondary neutrons from photo-neutron reactions on $\rm{^{208}Pb}$. Full size of the spectra up to 140 MeV as taken from ENDF database (a) and the low energy region of the same data (b)~\cite{Chadwick:2011xwu}.}
\label{fig:ENDF}
\end{center}
\end{figure}

The energy of the incident photon is expected to be distributed according to the partial cross sections. Neutrons are produced in the rest frame of the nucleus with an energy generated using the ENDF table (see Fig.~\ref{fig:ENDF}) and having an isotropic angular distribution. They are then  boosted to the laboratory frame either to positive or negative direction.

\section{Program flow}
\label{sec:ProgramFlow}

This section provides information on the flow of the program code.
\begin{itemize}
    \item Initialisation
    \begin{itemize}
        \item Data for TGraphs/TH1s as seen on Fig.~\ref{fig:xSectionXn} and Fig.~\ref{fig:xSectionNn} are loaded.
        \item The extrapolation fit shown in Fig.~\ref{fig:MultiExtrapol} is loaded and the branching-ratio map (Fig.~\ref{fig:MultiBRs}) is constructed.
        \item The hadronic interaction probability  as seen on Fig.~\ref{fig:HadronicProb} is constructed and stored as a TGraph.
        \item The nucleus break-up probability as a function of impact parameter is stored in a TGraph. By default up to 50 neutrons are constructed using Eq.~(\ref{eq:13}) and Eq.~(\ref{eq:17}) as partially shown in Fig.~\ref{fig:Breakup}.
        \item Photon fluxes are computed as a function of energy and stored as a TGraph. This includes the full unmodified flux, the denominator from Eq.~\ref{eq:10}, and fluxes for every combination of $i$,$j$ using the nucleus break-up probabilities from the previous step and Eq.~(\ref{eq:11}). The symmetry of the collision system is taken into account and the flux for any number of neutrons is stored as well. In total these are 1277 TGraphs for a maximum of 50 neutrons.
        \item The neutron emission spectra from the ENDF database are loaded either from the ASCII file itself or from a TH2 previously stored. 
        \item Run-time quality-assurance histograms are initialised.
        \item The  TTree event  with TParticle objects in a TClonesArray for the generated neutrons is initialised.
    \end{itemize}
\begin{figure}[t!]
\begin{center}
     \begin{subfigure}[b]{0.49\textwidth}
         \includegraphics[width=\textwidth]{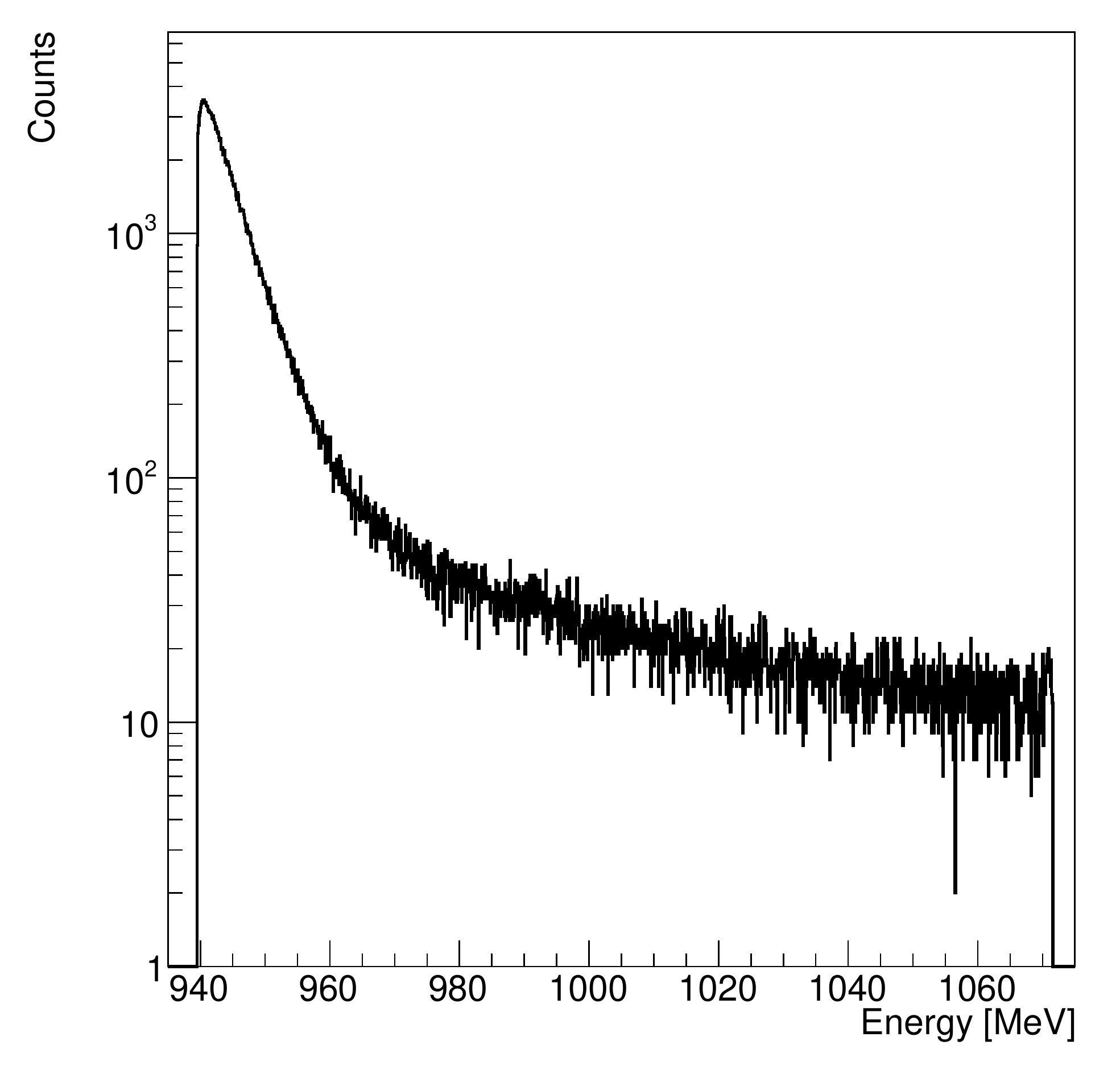}
         \caption{ }
         \label{fig:Gen1_a}
     \end{subfigure}
     \begin{subfigure}[b]{0.49\textwidth}
         \includegraphics[width=\textwidth]{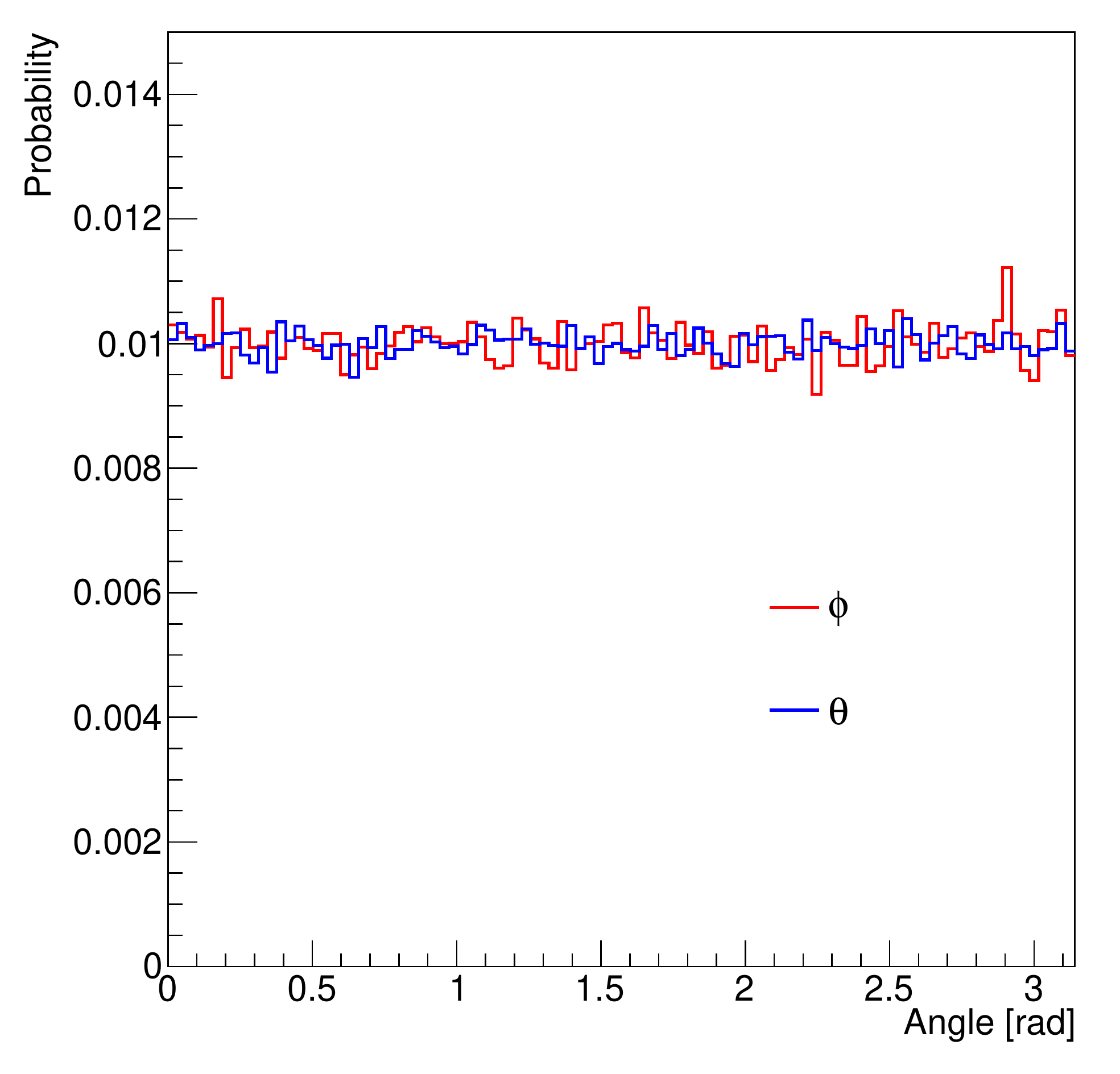}
         \caption{ }
         \label{fig:Gen1_b}
     \end{subfigure}
\caption{(Colour online) Properties of the generated neutrons  in the rest frame of the emitting nucleus. Total energy (a) and isotropic angular distributions (b) for the polar (blue) and azimuth (red) angles.}
\label{fig:Gen1}
\end{center}
\end{figure}

\begin{figure}[t!]
\begin{center}
     \begin{subfigure}[b]{0.49\textwidth}
         \includegraphics[width=\textwidth]{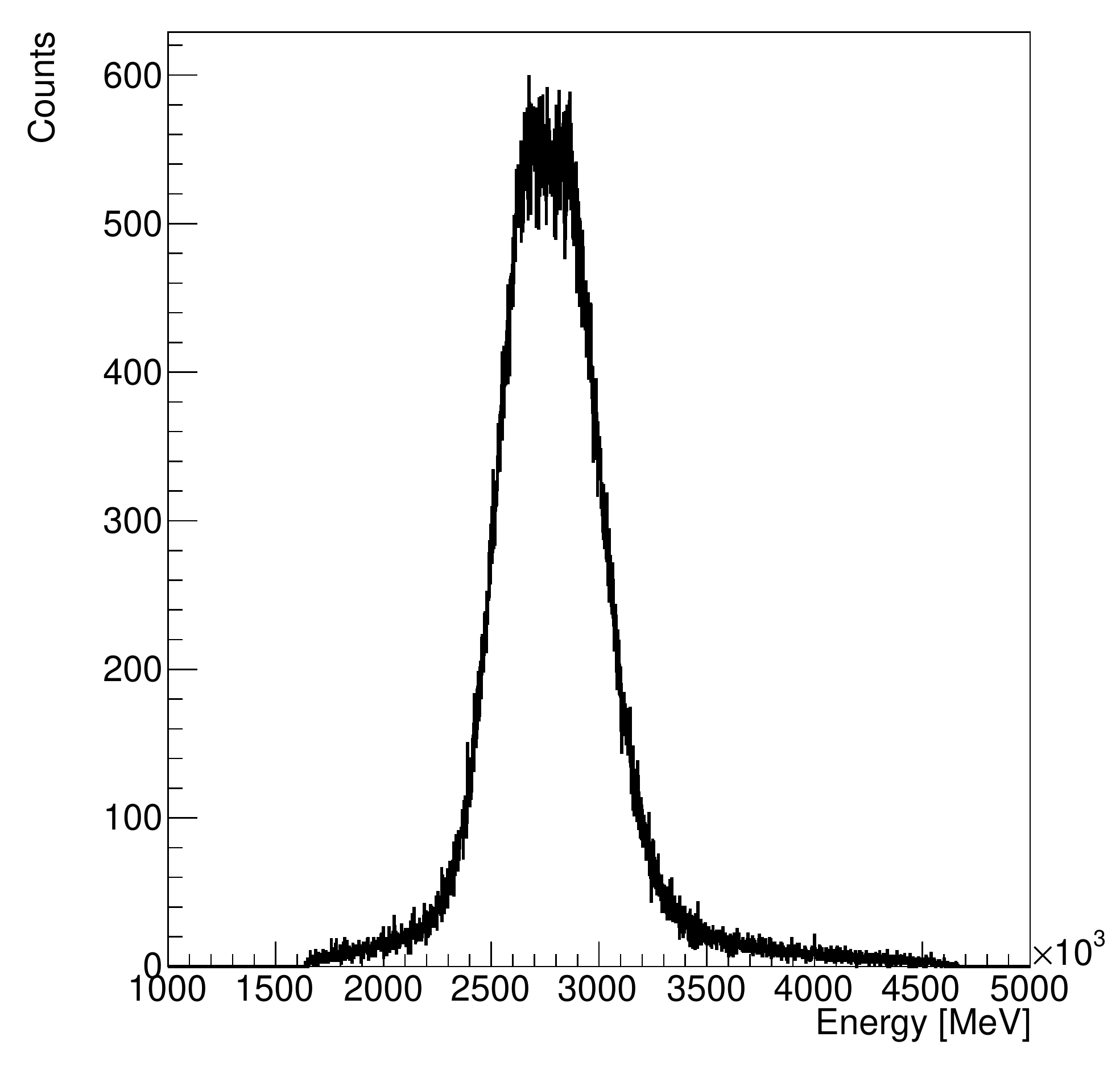}
         \caption{ }
         \label{fig:Gen2_a}
     \end{subfigure}
     \begin{subfigure}[b]{0.49\textwidth}
         \includegraphics[width=\textwidth]{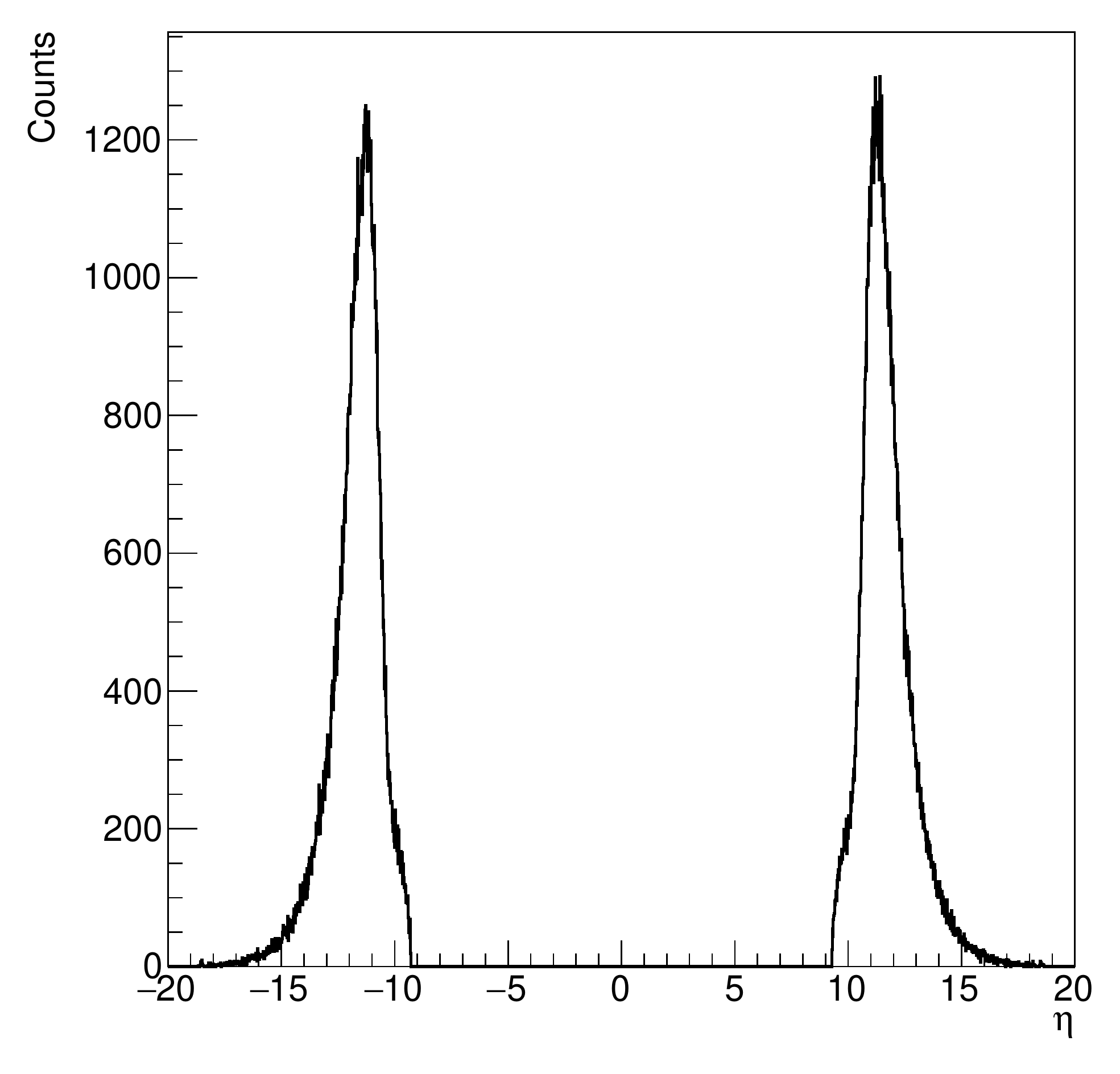}
         \caption{ }
         \label{fig:Gen2_b}
     \end{subfigure}
\caption{Properties of the generated neutrons in the LHC frame for Run 2 energies. Total energy (a) and pseudorapidity $\eta$ (b).}
\label{fig:Gen3}
\end{center}
\end{figure}
    \item Event generation
    \begin{itemize}
        \item An external entity must provide the photon energy participating in the process $P$ in an event-by-event basis. Two possibilities are currently implemented in the generator. It can either run together with the STARlight Monte Carlo program~\cite{Klein:2016yzr} or use theory curves for photonuclear cross sections. The generator can also be  easily interfaced to any external software package.
        \item The triangular map of break-up probabilities per number of neutrons on either side is constructed for the current photon energy. 
        \item The number of neutrons to create on each side is defined as a random variable  distributed according to the map from the previous step.
    \end{itemize}
    \item Particle generation
    \begin{itemize}
        \item The emission spectra from the ENDF database are used here. The break-up photon energy is distributed according to the partial cross section. Once the multiplicity of neutrons per side is known,  the break-up photon energy is generated.
        \item Using the emission spectra the bin corresponding to the energy of break-up photon is found. As shown on Fig.~\ref{fig:ENDF} the spectra goes only up to 140 MeV. For energies higher than 140 MeV the last bin of the spectrum is used.
        \item Neutrons are created in the rest frame of the nucleus with the kinetic energy distributed according to the one-dimensional slice of the emission spectra from the previous step. The total momentum is computed from the kinetic energy, and the angular distributions are generated to be isotropic. These kinematic distributions are shown in  Fig.~\ref{fig:Gen1}. 
        \item The neutrons are boosted into the laboratory frame. The energy and pseudorapidity distributions  of the boosted neutrons are shown in Fig.~\ref{fig:Gen3}.
        \item TParticle objects are created on the TClonesArray and stored in the event TTree. At this step the units are changed from MeV to GeV.
    \end{itemize}
\end{itemize}

\section{Example of possible applications}
\label{sec:Examples}

\subsection{Results from running as STARlight afterburner}
The STARlight generator~\cite{Klein:2016yzr} is currently the most popular Monte Carlo program for UPC studies. For this reason we have used it here to provide the example. However, we expect that other MC such as PYTHIA~\cite{Sjostrand:2007gs} or SuperChic~\cite{Harland-Lang:2018iur} can be easily coupled to Noon. 

An interface to STARlight is provided through the ROOT class  TStarlight, so that to each  event one can add  the neutrons produced  by the \Noon\ generator. It is then trivial to pass these neutrons (along with the particles produced by STARlight) through the detailed simulation of the experiments.

\begin{figure}[htbp]
\begin{center}
\begin{subfigure}[b]{0.49\textwidth}
         \includegraphics[width=\textwidth]{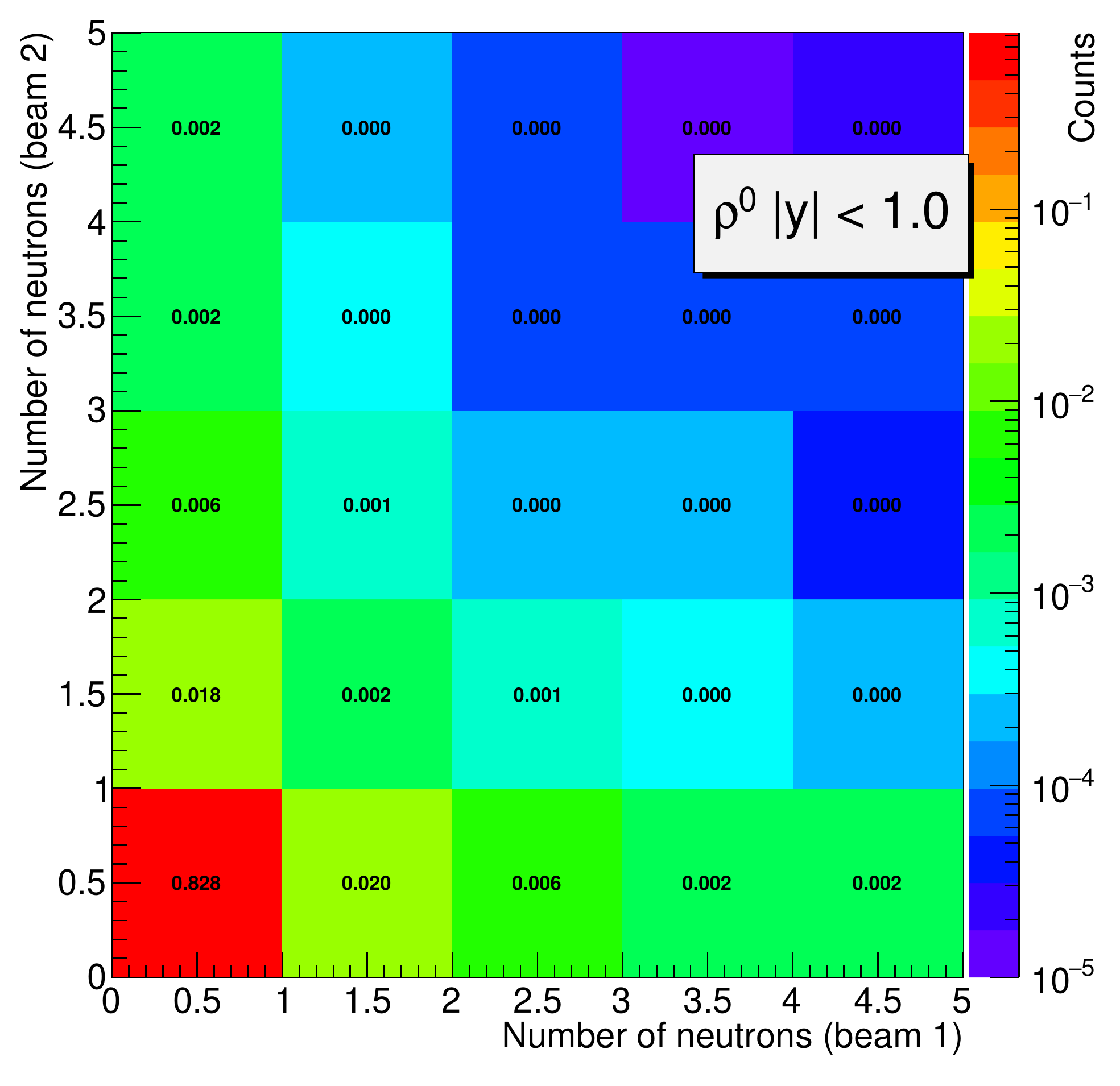}
         \caption{ }
         \label{fig:ResultSL_a}
     \end{subfigure}
\begin{subfigure}[b]{0.49\textwidth}
         \includegraphics[width=\textwidth]{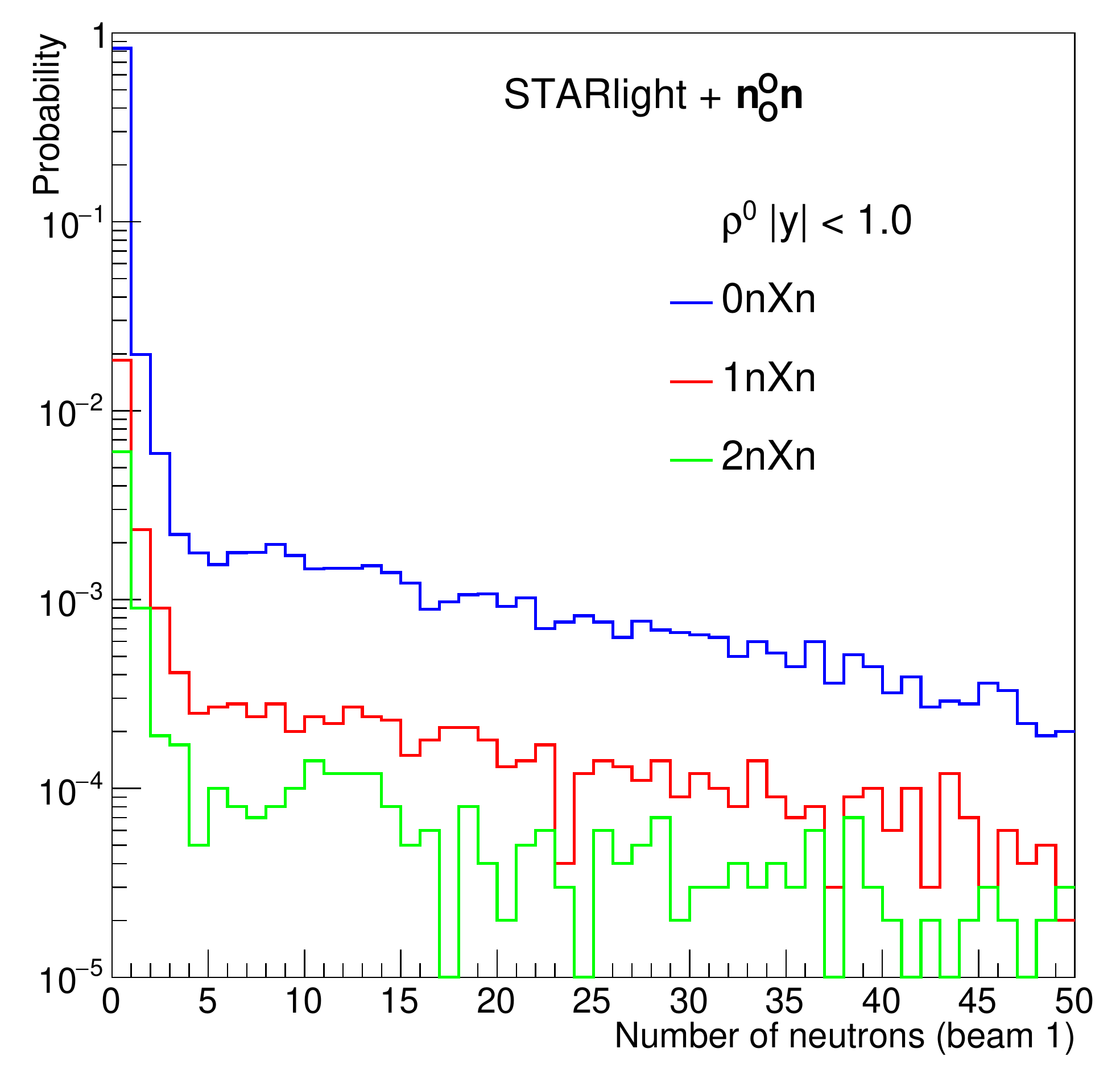}
         \caption{ }
         \label{fig:ResultSL_b}
     \end{subfigure}
\caption{(Colour online) Normalised neutron multiplicity distributions for coherent production of $\rho^{0}$ $(a)$ at the LHC energies of Run 2 in Pb--Pb UPC for events generated with the STARlight program. The vector mesons were generated at mid-rapidity $|y|<1.0$. The blue (red, green) line represent events with no (one, two) neutrons in one side an any number (larger than zero, Xn) of neutrons in the other side. }
\label{fig:ResultSL}
\end{center}
\end{figure}

As an example, Fig.~\ref{fig:ResultSL}  shows the neutron multiplicity distributions for the coherent production of  $\rho^{0}$  at mid-rapidity for Run 2 LHC energies for events generated by STARlight.  

\subsection{Results from running with theory input}
\label{sec:hotspot}

The neutron generator \Noon\ can also use  theory predictions as a function of rapidity for the cross section of the coherent photonuclear production of a vector meson, together with the mass distribution of the vector meson (see Eq.~(\ref{eq:8}), to produce neutron multiplicities in a selected rapidity range. 

\begin{figure}[htbp]
\begin{center}
\includegraphics[width=0.49\textwidth]{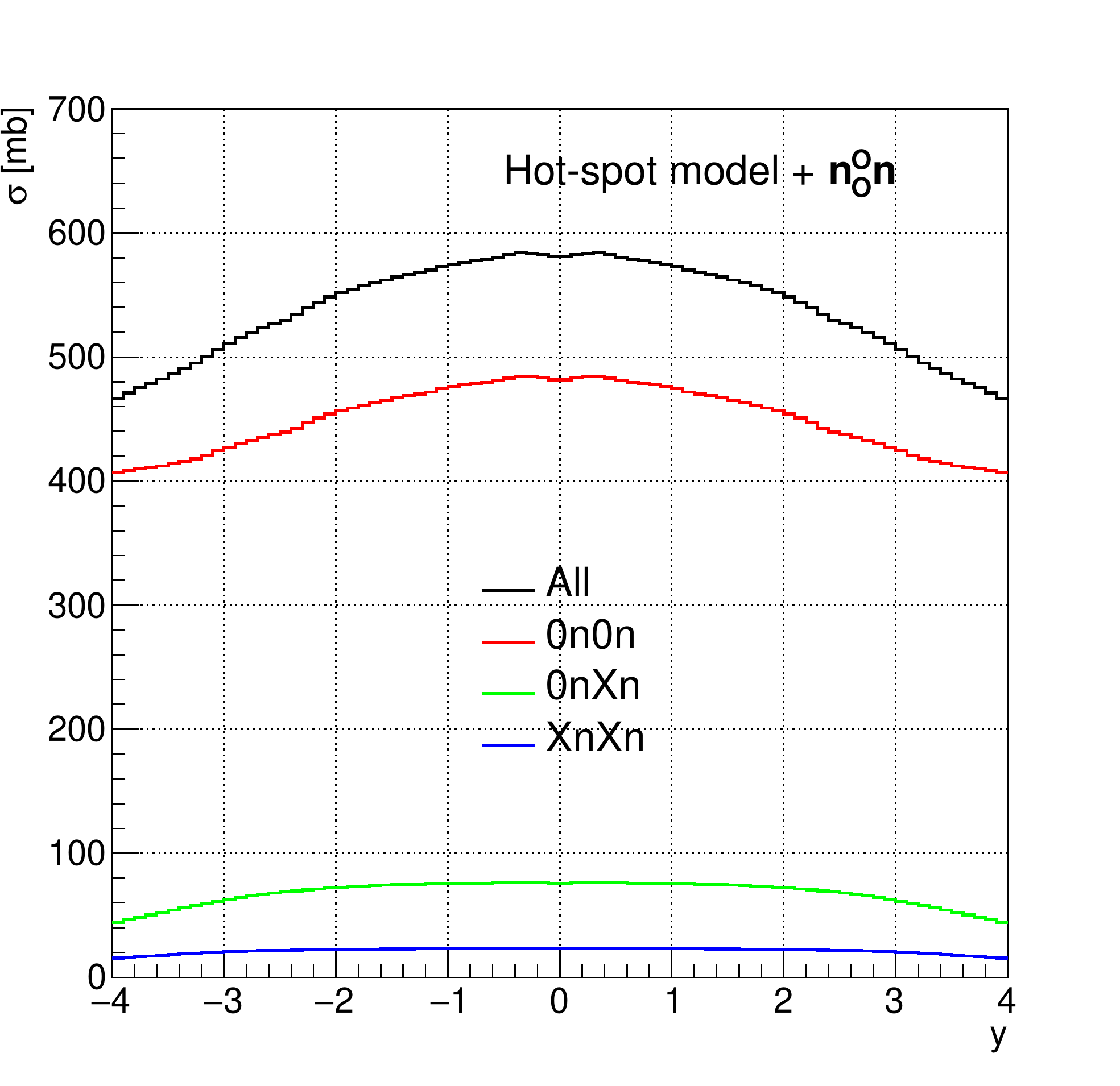}
\includegraphics[width=0.49\textwidth]{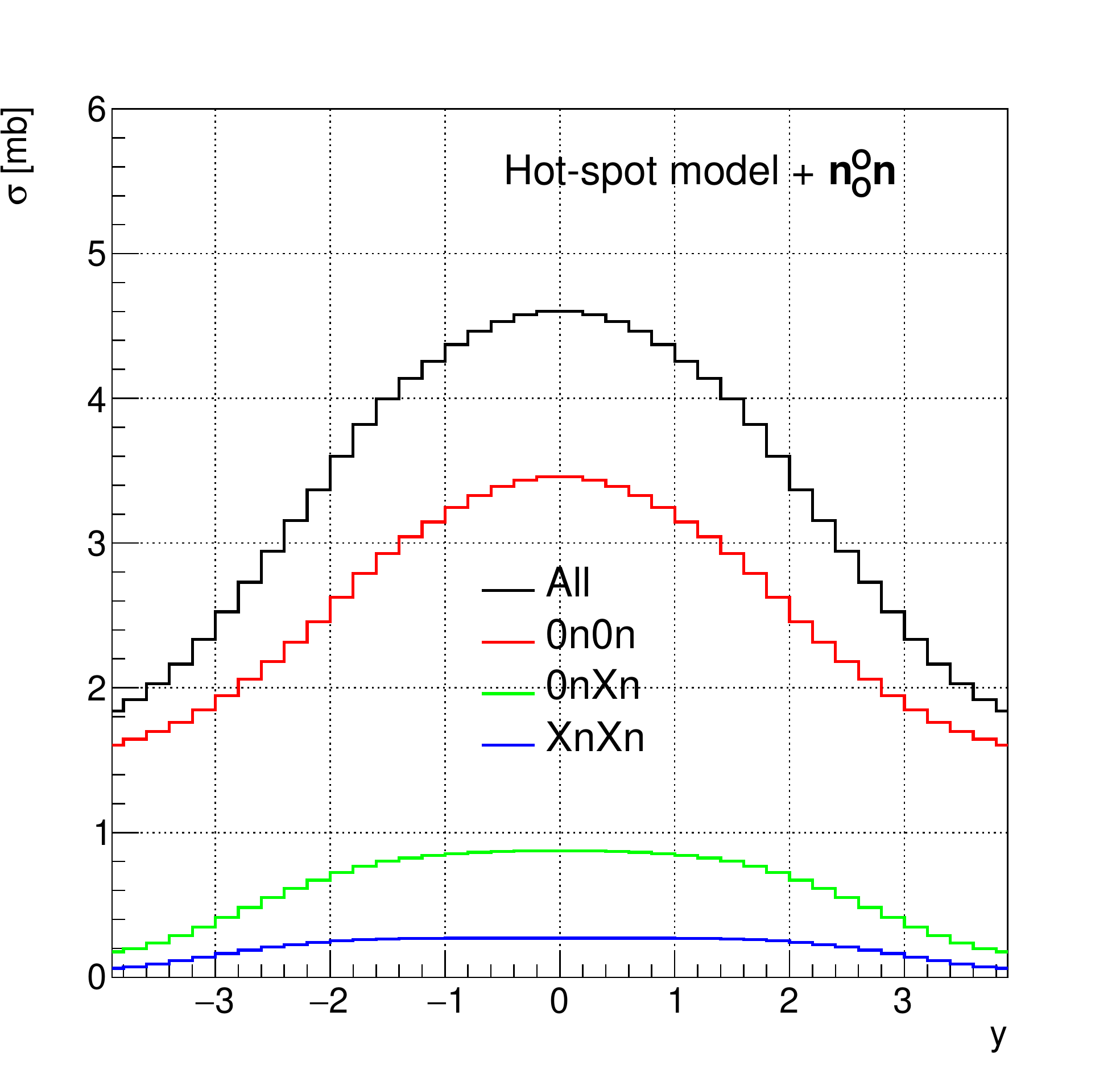}
\caption{(Colour online) Expansion of the predictions for  coherent photonuclear production of $\rho^{0}$ (left) and ${\rm J}/\psi$ (right) as function of rapidity into different classes of neutrons at beam rapidities. The energy-dependent hot-spot model~\cite{Cepila:2016uku,Cepila:2018zky,Cepila:2017nef} is used as input to the \Noon\ program.}
\label{fig:ResultGC}
\end{center}
\end{figure}

As an example, the predictions of the energy-dependent hot-spot model~\cite{Cepila:2016uku} applied to the coherent photonuclear production of a $\rho^{0}$ and ${\rm J}/\psi$ vector meson~\cite{Cepila:2018zky,Cepila:2017nef} were taken as an input. These cross sections are total, in the sense that are independent of the potential  emission of neutrons at beam rapidities. Using the \Noon\ program the predictions of this model were separated into the contributions from events without neutrons (0n0n) with neutrons in  only one side (0nXn+Xn0n) and with neutrons in both sides (XnXn). The resulting cross sections are shown in Fig.~\ref{fig:ResultGC}.

\FloatBarrier
\section{Conclusions and outlook}
\label{sec:Conclusions}

We have presented the \Noon\ Monte Carlo program that fills an important void for studies of photon-induced processes at the LHC. \Noon\ produces in an event-by-event basis neutrons originating from electromagnetic dissociation interactions between the colliding heavy nuclei at facilities like RHIC and the LHC. 

The input to \Noon\ can be either the STARlight MC program or theoretical predictions for hard reactions in photonuclear processes, like the energy-dependent hot-spot model discussed above. The output of \Noon\ is easily interfaced to the simulation program of the collaborations at RHIC and the LHC. 

Although we have focused the examples on coherent vector meson production,  \Noon\ can also be very promising for studies of other photon-induced processes such as jets in UPC~\cite{ATLAS:2017kwa} and light-by-light scattering~\cite{Aaboud:2017bwk,Aad:2019ock,Sirunyan:2018fhl}.

\section*{Acknowledgements}
This work was partially supported by grants 18-07880S of the Czech Science Foundation and LTC17038 of the INTER-EXCELLENCE program at the Ministry of Education, Youth and Sports of the Czech Republic, and the U.S. Department of Energy, Office of Science, Nuclear Physics under award number FED30210.


\bibliographystyle{elsarticle-num}
\bibliography{Noon}

\begin{thebibliography}{10}
\expandafter\ifx\csname url\endcsname\relax
  \def\url#1{\texttt{#1}}\fi
\expandafter\ifx\csname urlprefix\endcsname\relax\def\urlprefix{URL }\fi
\expandafter\ifx\csname href\endcsname\relax
  \def\href#1#2{#2} \def\path#1{#1}\fi

\bibitem{Fermi:1924tc}
E.~Fermi, {On the Theory of the impact between atoms and electrically charged
  particles}, Z.Phys. 29 (1924) 315--327.
\newblock \href {http://dx.doi.org/10.1007/BF03184853}
  {\path{doi:10.1007/BF03184853}}.

\bibitem{Fermi:1925fq}
E.~Fermi, {On the theory of collisions between atoms and electrically charged
  particles}, Nuovo Cim. 2 (1925) 143--158.
\newblock \href {http://arxiv.org/abs/hep-th/0205086}
  {\path{arXiv:hep-th/0205086}}, \href {http://dx.doi.org/10.1007/BF02961914}
  {\path{doi:10.1007/BF02961914}}.

\bibitem{Krauss:1997vr}
F.~Krauss, M.~Greiner, G.~Soff, {Photon and gluon induced processes in
  relativistic heavy ion collisions}, Prog. Part. Nucl. Phys. 39 (1997)
  503--564.
\newblock \href {http://dx.doi.org/10.1016/S0146-6410(97)00049-5}
  {\path{doi:10.1016/S0146-6410(97)00049-5}}.

\bibitem{Baur:2001jj}
G.~Baur, K.~Hencken, D.~Trautmann, S.~Sadovsky, Y.~Kharlov, {Coherent gamma
  gamma and gamma-A interactions in very peripheral collisions at relativistic
  ion colliders}, Phys.Rept. 364 (2002) 359--450.
\newblock \href {http://arxiv.org/abs/hep-ph/0112211}
  {\path{arXiv:hep-ph/0112211}}, \href
  {http://dx.doi.org/10.1016/S0370-1573(01)00101-6}
  {\path{doi:10.1016/S0370-1573(01)00101-6}}.

\bibitem{Bertulani:2005ru}
C.~A. Bertulani, S.~R. Klein, J.~Nystrand, {Physics of ultra-peripheral nuclear
  collisions}, Ann. Rev. Nucl. Part. Sci. 55 (2005) 271--310.
\newblock \href {http://arxiv.org/abs/nucl-ex/0502005}
  {\path{arXiv:nucl-ex/0502005}}, \href
  {http://dx.doi.org/10.1146/annurev.nucl.55.090704.151526}
  {\path{doi:10.1146/annurev.nucl.55.090704.151526}}.

\bibitem{Baltz:2007kq}
A.~J. Baltz, {The Physics of Ultraperipheral Collisions at the LHC}, Phys.
  Rept. 458 (2008) 1--171.
\newblock \href {http://arxiv.org/abs/0706.3356} {\path{arXiv:0706.3356}},
  \href {http://dx.doi.org/10.1016/j.physrep.2007.12.001}
  {\path{doi:10.1016/j.physrep.2007.12.001}}.

\bibitem{Contreras:2015dqa}
J.~G. Contreras, J.~D. Tapia~Takaki, {Ultra-peripheral heavy-ion collisions at
  the LHC}, Int. J. Mod. Phys. A30 (2015) 1542012.
\newblock \href {http://dx.doi.org/10.1142/S0217751X15420129}
  {\path{doi:10.1142/S0217751X15420129}}.

\bibitem{Ryskin:1992ui}
M.~G. Ryskin, {Diffractive J / psi electroproduction in LLA QCD}, Z. Phys. C57
  (1993) 89--92.
\newblock \href {http://dx.doi.org/10.1007/BF01555742}
  {\path{doi:10.1007/BF01555742}}.

\bibitem{Agakishiev:2011me}
G.~Agakishiev, et~al., {$\rho^{0}$ Photoproduction in AuAu Collisions at
  $\sqrt{s_{NN}}$=62.4 GeV with STAR}, Phys. Rev. C85 (2012) 014910.
\newblock \href {http://arxiv.org/abs/1107.4630} {\path{arXiv:1107.4630}},
  \href {http://dx.doi.org/10.1103/PhysRevC.85.014910}
  {\path{doi:10.1103/PhysRevC.85.014910}}.

\bibitem{Adler:2002sc}
C.~Adler, et~al., {Coherent rho0 production in ultraperipheral heavy ion
  collisions}, Phys. Rev. Lett. 89 (2002) 272302.
\newblock \href {http://arxiv.org/abs/nucl-ex/0206004}
  {\path{arXiv:nucl-ex/0206004}}, \href
  {http://dx.doi.org/10.1103/PhysRevLett.89.272302}
  {\path{doi:10.1103/PhysRevLett.89.272302}}.

\bibitem{Abelev:2007nb}
B.~I. Abelev, et~al., {$\rho^0$ photoproduction in ultraperipheral relativistic
  heavy ion collisions at $\sqrt{s_{NN}}$ = 200 GeV}, Phys. Rev. C77 (2008)
  034910.
\newblock \href {http://arxiv.org/abs/0712.3320} {\path{arXiv:0712.3320}},
  \href {http://dx.doi.org/10.1103/PhysRevC.77.034910}
  {\path{doi:10.1103/PhysRevC.77.034910}}.

\bibitem{Adamczyk:2017vfu}
L.~Adamczyk, et~al., {Coherent diffractive photoproduction of ?0 mesons on gold
  nuclei at 200 GeV/nucleon-pair at the Relativistic Heavy Ion Collider}, Phys.
  Rev. C96~(5) (2017) 054904.
\newblock \href {http://arxiv.org/abs/1702.07705} {\path{arXiv:1702.07705}},
  \href {http://dx.doi.org/10.1103/PhysRevC.96.054904}
  {\path{doi:10.1103/PhysRevC.96.054904}}.

\bibitem{Afanasiev:2009hy}
S.~Afanasiev, et~al., {Photoproduction of J/psi and of high mass e+e- in
  ultra-peripheral Au+Au collisions at s**(1/2) = 200-GeV}, Phys. Lett. B679
  (2009) 321--329.
\newblock \href {http://arxiv.org/abs/0903.2041} {\path{arXiv:0903.2041}},
  \href {http://dx.doi.org/10.1016/j.physletb.2009.07.061}
  {\path{doi:10.1016/j.physletb.2009.07.061}}.

\bibitem{Adam:2015gsa}
J.~Adam, et~al., {Coherent $\rho^{0}$ photoproduction in ultra-peripheral Pb-Pb
  collisions at $ \sqrt{s_{\mathrm{NN}}}=2.76 $ TeV}, JHEP 09 (2015) 095.
\newblock \href {http://arxiv.org/abs/1503.09177} {\path{arXiv:1503.09177}},
  \href {http://dx.doi.org/10.1007/JHEP09(2015)095}
  {\path{doi:10.1007/JHEP09(2015)095}}.

\bibitem{Sirunyan:2019nog}
A.~M. Sirunyan, et~al., {Measurement of exclusive $\rho(770)^0$ photoproduction
  in ultraperipheral pPb collisions at $\sqrt{s_\mathrm{NN}} =$ 5.02 TeV}\href
  {http://arxiv.org/abs/1902.01339} {\path{arXiv:1902.01339}}.

\bibitem{Abelev:2012ba}
B.~Abelev, et~al., {Coherent $J/\psi$ photoproduction in ultra-peripheral Pb-Pb
  collisions at $\sqrt{s_{NN}} = 2.76$ TeV}, Phys. Lett. B718 (2013)
  1273--1283.
\newblock \href {http://arxiv.org/abs/1209.3715} {\path{arXiv:1209.3715}},
  \href {http://dx.doi.org/10.1016/j.physletb.2012.11.059}
  {\path{doi:10.1016/j.physletb.2012.11.059}}.

\bibitem{Abbas:2013oua}
E.~Abbas, et~al., {Charmonium and $e^+e^-$ pair photoproduction at mid-rapidity
  in ultra-peripheral Pb-Pb collisions at $\sqrt{s_{\rm NN}}$=2.76 TeV}, Eur.
  Phys. J. C73~(11) (2013) 2617.
\newblock \href {http://arxiv.org/abs/1305.1467} {\path{arXiv:1305.1467}},
  \href {http://dx.doi.org/10.1140/epjc/s10052-013-2617-1}
  {\path{doi:10.1140/epjc/s10052-013-2617-1}}.

\bibitem{Khachatryan:2016qhq}
V.~Khachatryan, et~al., {Coherent $J/\psi$ photoproduction in ultra-peripheral
  PbPb collisions at $\sqrt {s_{NN}} =$ 2.76 TeV with the CMS experiment},
  Phys. Lett. B772 (2017) 489--511.
\newblock \href {http://arxiv.org/abs/1605.06966} {\path{arXiv:1605.06966}},
  \href {http://dx.doi.org/10.1016/j.physletb.2017.07.001}
  {\path{doi:10.1016/j.physletb.2017.07.001}}.

\bibitem{Acharya:2018jua}
S.~Acharya, et~al., {Energy dependence of exclusive $\mathrm {J}/\psi $
  photoproduction off protons in ultra-peripheral p–Pb collisions at
  $\sqrt{s_{\mathrm {\scriptscriptstyle NN}}} = 5.02$ TeV}, Eur. Phys. J.
  C79~(5) (2019) 402.
\newblock \href {http://arxiv.org/abs/1809.03235} {\path{arXiv:1809.03235}},
  \href {http://dx.doi.org/10.1140/epjc/s10052-019-6816-2}
  {\path{doi:10.1140/epjc/s10052-019-6816-2}}.

\bibitem{Acharya:2019vlb}
S.~Acharya, et~al., {Coherent J/$\psi$ photoproduction at forward rapidity in
  ultra-peripheral Pb-Pb collisions at $\sqrt{s_{\rm{NN}}}=5.02$ TeV}\href
  {http://arxiv.org/abs/1904.06272} {\path{arXiv:1904.06272}}.

\bibitem{Adam:2015sia}
J.~Adam, et~al., {Coherent $\psi$(2S) photo-production in ultra-peripheral Pb
  Pb collisions at $\sqrt{s}_{\rm NN}$ = 2.76 TeV}, Phys. Lett. B751 (2015)
  358--370.
\newblock \href {http://arxiv.org/abs/1508.05076} {\path{arXiv:1508.05076}},
  \href {http://dx.doi.org/10.1016/j.physletb.2015.10.040}
  {\path{doi:10.1016/j.physletb.2015.10.040}}.

\bibitem{Sirunyan:2018sav}
A.~M. Sirunyan, et~al., {Measurement of exclusive $\Upsilon$ photoproduction
  from protons in pPb collisions at $\sqrt{s_\mathrm{NN}} =$ 5.02 TeV}, Eur.
  Phys. J. C79~(3) (2019) 277.
\newblock \href {http://arxiv.org/abs/1809.11080} {\path{arXiv:1809.11080}},
  \href {http://dx.doi.org/10.1140/epjc/s10052-019-6774-8}
  {\path{doi:10.1140/epjc/s10052-019-6774-8}}.

\bibitem{ALICE:2012aa}
B.~Abelev, et~al., {Measurement of the Cross Section for Electromagnetic
  Dissociation with Neutron Emission in Pb-Pb Collisions at $\sqrt{s_{NN}}$ =
  2.76 TeV}, Phys. Rev. Lett. 109 (2012) 252302.
\newblock \href {http://arxiv.org/abs/1203.2436} {\path{arXiv:1203.2436}},
  \href {http://dx.doi.org/10.1103/PhysRevLett.109.252302}
  {\path{doi:10.1103/PhysRevLett.109.252302}}.

\bibitem{Pshenichnov:2001qd}
I.~A. Pshenichnov, J.~P. Bondorf, I.~N. Mishustin, A.~Ventura, S.~Masetti,
  {Mutual heavy ion dissociation in peripheral collisions at ultrarelativistic
  energies}, Phys. Rev. C64 (2001) 024903.
\newblock \href {http://arxiv.org/abs/nucl-th/0101035}
  {\path{arXiv:nucl-th/0101035}}, \href
  {http://dx.doi.org/10.1103/PhysRevC.64.024903}
  {\path{doi:10.1103/PhysRevC.64.024903}}.

\bibitem{Pshenichnov:2011zz}
I.~A. Pshenichnov, {Electromagnetic excitation and fragmentation of
  ultrarelativistic nuclei}, Phys. Part. Nucl. 42 (2011) 215--250.
\newblock \href {http://dx.doi.org/10.1134/S1063779611020067}
  {\path{doi:10.1134/S1063779611020067}}.

\bibitem{Baltz:2002pp}
A.~J. Baltz, S.~R. Klein, J.~Nystrand, {Coherent vector meson photoproduction
  with nuclear breakup in relativistic heavy ion collisions}, Phys. Rev. Lett.
  89 (2002) 012301.
\newblock \href {http://arxiv.org/abs/nucl-th/0205031}
  {\path{arXiv:nucl-th/0205031}}, \href
  {http://dx.doi.org/10.1103/PhysRevLett.89.012301}
  {\path{doi:10.1103/PhysRevLett.89.012301}}.

\bibitem{Guzey:2013jaa}
V.~Guzey, M.~Strikman, M.~Zhalov, {Disentangling coherent and incoherent
  quasielastic $J/\psi$ photoproduction on nuclei by neutron tagging in
  ultraperipheral ion collisions at the LHC}, Eur. Phys. J. C74~(7) (2014)
  2942.
\newblock \href {http://arxiv.org/abs/1312.6486} {\path{arXiv:1312.6486}},
  \href {http://dx.doi.org/10.1140/epjc/s10052-014-2942-z}
  {\path{doi:10.1140/epjc/s10052-014-2942-z}}.

\bibitem{Guzey:2016piu}
V.~Guzey, E.~Kryshen, M.~Zhalov, {Coherent photoproduction of vector mesons in
  ultraperipheral heavy ion collisions: Update for run 2 at the CERN Large
  Hadron Collider}, Phys. Rev. C93~(5) (2016) 055206.
\newblock \href {http://arxiv.org/abs/1602.01456} {\path{arXiv:1602.01456}},
  \href {http://dx.doi.org/10.1103/PhysRevC.93.055206}
  {\path{doi:10.1103/PhysRevC.93.055206}}.

\bibitem{Klein:2016yzr}
S.~R. Klein, J.~Nystrand, J.~Seger, Y.~Gorbunov, J.~Butterworth, {STARlight: A
  Monte Carlo simulation program for ultra-peripheral collisions of
  relativistic ions}, Comput. Phys. Commun. 212 (2017) 258--268.
\newblock \href {http://arxiv.org/abs/1607.03838} {\path{arXiv:1607.03838}},
  \href {http://dx.doi.org/10.1016/j.cpc.2016.10.016}
  {\path{doi:10.1016/j.cpc.2016.10.016}}.

\bibitem{Veyssiere:1970ztg}
A.~Veyssiere, H.~Beil, R.~Bergere, P.~Carlos, A.~Lepretre, {Photoneutron cross
  sections of 208 Pb and 197 Au}, Nucl. Phys. A159 (1970) 561--576.
\newblock \href {http://dx.doi.org/10.1016/0375-9474(70)90727-X}
  {\path{doi:10.1016/0375-9474(70)90727-X}}.

\bibitem{Lepretre:1981tf}
A.~Lepretre, H.~Beil, R.~Bergere, P.~Carlos, J.~Fagot, A.~De~Miniac,
  A.~Veyssiere, {Measurements of the Total Photonuclear Cross-sections From
  30-{MeV} to 140-{MeV} for {SN}, Ce, Ta, Pb and U Nuclei}, Nucl. Phys. A367
  (1981) 237--268.
\newblock \href {http://dx.doi.org/10.1016/0375-9474(81)90516-9}
  {\path{doi:10.1016/0375-9474(81)90516-9}}.

\bibitem{Carlos:1985te}
P.~Carlos, H.~Beil, R.~Bergere, J.~Fagot, A.~Lepretre, A.~de~Miniac,
  A.~Veyssiere, {TOTAL PHOTONUCLEAR ABSORPTION CROSS-SECTION FOR PB AND FOR
  HEAVY NUCLEI IN THE DELTA RESONANCE REGION}, Nucl. Phys. A431 (1984)
  573--592.
\newblock \href {http://dx.doi.org/10.1016/0375-9474(84)90269-0}
  {\path{doi:10.1016/0375-9474(84)90269-0}}.

\bibitem{Armstrong:1971ns}
T.~A. Armstrong, et~al., {Total hadronic cross-section of gamma rays in
  hydrogen in the energy range 0.265-GeV to 4.215-GeV}, Phys. Rev. D5 (1972)
  1640--1652.
\newblock \href {http://dx.doi.org/10.1103/PhysRevD.5.1640}
  {\path{doi:10.1103/PhysRevD.5.1640}}.

\bibitem{Armstrong:1972sa}
T.~A. Armstrong, et~al., {The total photon deuteron hadronic cross-section in
  the energy range 0.265-4.215 gev}, Nucl. Phys. B41 (1972) 445--473.
\newblock \href {http://dx.doi.org/10.1016/0550-3213(72)90403-8}
  {\path{doi:10.1016/0550-3213(72)90403-8}}.

\bibitem{Michalowski:1977eg}
S.~Michalowski, D.~Andrews, J.~Eickmeyer, T.~Gentile, N.~B. Mistry, R.~Talman,
  K.~Ueno, {Experimental Study of Nuclear Shadowing in Photoproduction}, Phys.
  Rev. Lett. 39 (1977) 737--740.
\newblock \href {http://dx.doi.org/10.1103/PhysRevLett.39.737}
  {\path{doi:10.1103/PhysRevLett.39.737}}.

\bibitem{Caldwell:1973bu}
D.~O. Caldwell, V.~B. Elings, W.~P. Hesse, R.~J. Morrison, F.~V. Murphy, D.~E.
  Yount, {Total Hadronic Photoabsorption Cross-Sections on Hydrogen and Complex
  Nuclei from 4-GeV to 18-GeV}, Phys. Rev. D7 (1973) 1362.
\newblock \href {http://dx.doi.org/10.1103/PhysRevD.7.1362}
  {\path{doi:10.1103/PhysRevD.7.1362}}.

\bibitem{Caldwell:1978ik}
D.~O. Caldwell, et~al., {Measurement of Shadowing in Photon - Nucleus Total
  Cross-sections From 20-{GeV} to 185-{GeV}}, Phys. Rev. Lett. 42 (1979) 553.
\newblock \href {http://dx.doi.org/10.1103/PhysRevLett.42.553}
  {\path{doi:10.1103/PhysRevLett.42.553}}.

\bibitem{Berman:1987zz}
B.~L. Berman, R.~E. Pywell, S.~S. Dietrich, M.~N. Thompson, K.~G. McNeill,
  J.~W. Jury, {Absolute photoneutron cross sections for Zr, I, Pr, Au, and Pb},
  Phys. Rev. C36 (1987) 1286--1292.
\newblock \href {http://dx.doi.org/10.1103/PhysRevC.36.1286}
  {\path{doi:10.1103/PhysRevC.36.1286}}.

\bibitem{Baltz:1996as}
A.~J. Baltz, M.~J. Rhoades-Brown, J.~Weneser, {Heavy ion partial beam lifetimes
  due to Coulomb induced processes}, Phys. Rev. E54 (1996) 4233--4239.
\newblock \href {http://dx.doi.org/10.1103/PhysRevE.54.4233}
  {\path{doi:10.1103/PhysRevE.54.4233}}.

\bibitem{Baltz:1998ex}
A.~J. Baltz, C.~Chasman, S.~N. White, {Correlated forward - backward
  dissociation and neutron spectra as luminosity monitor in heavy ion
  colliders}, Nucl. Instrum. Meth. A417 (1998) 1--8.
\newblock \href {http://arxiv.org/abs/nucl-ex/9801002}
  {\path{arXiv:nucl-ex/9801002}}, \href
  {http://dx.doi.org/10.1016/S0168-9002(98)00575-0}
  {\path{doi:10.1016/S0168-9002(98)00575-0}}.

\bibitem{Lepretre:1982xs}
A.~Lepretre, H.~Beil, R.~Bergere, P.~Carlos, J.~Fagot, A.~Veyssiere,
  I.~Halpern, {ANALYSIS OF NEUTRON MULTIPLICITIES IN PHOTONUCLEAR REACTIONS
  FROM 30-MEV TO 140-MEV IN HEAVY ELEMENTS}, Nucl. Phys. A390 (1982) 221--239.
\newblock \href {http://dx.doi.org/10.1016/0375-9474(82)90159-2}
  {\path{doi:10.1016/0375-9474(82)90159-2}}.

\bibitem{ArrudaNeto:1977uf}
J.~D.~T. Arruda-Neto, S.~Simionatto, V.~P. Likhachev, F.~Garcia, J.~Mesa,
  A.~Deppman, O.~Rodriguez, F.~Guzman, {Photoneutron multiplicities of
  preactinide nuclei at energies above the pion threshold}, Nucl. Phys. A638
  (1998) 701--713.
\newblock \href {http://dx.doi.org/10.1016/S0375-9474(98)00215-2}
  {\path{doi:10.1016/S0375-9474(98)00215-2}}.

\bibitem{Chadwick:2011xwu}
M.~B. Chadwick, et~al., {ENDF/B-VII.1 Nuclear Data for Science and Technology:
  Cross Sections, Covariances, Fission Product Yields and Decay Data}, Nucl.
  Data Sheets 112~(12) (2011) 2887--2996.
\newblock \href {http://dx.doi.org/10.1016/j.nds.2011.11.002}
  {\path{doi:10.1016/j.nds.2011.11.002}}.

\bibitem{osti_981813}
M.~Herman, M.~of~the Cross Sections Evaluation Working~Group, Endf-6 formats
  manual data formats and procedures for the evaluated nuclear data file
  endf/b-vi and endf/b-vii\href {http://dx.doi.org/10.2172/981813}
  {\path{doi:10.2172/981813}}.

\bibitem{Sjostrand:2007gs}
T.~Sjostrand, S.~Mrenna, P.~Z. Skands, {A Brief Introduction to PYTHIA 8.1},
  Comput. Phys. Commun. 178 (2008) 852--867.
\newblock \href {http://arxiv.org/abs/0710.3820} {\path{arXiv:0710.3820}},
  \href {http://dx.doi.org/10.1016/j.cpc.2008.01.036}
  {\path{doi:10.1016/j.cpc.2008.01.036}}.

\bibitem{Harland-Lang:2018iur}
L.~A. Harland-Lang, V.~A. Khoze, M.~G. Ryskin, {Exclusive LHC physics with
  heavy ions: SuperChic 3}, Eur. Phys. J. C79~(1) (2019) 39.
\newblock \href {http://arxiv.org/abs/1810.06567} {\path{arXiv:1810.06567}},
  \href {http://dx.doi.org/10.1140/epjc/s10052-018-6530-5}
  {\path{doi:10.1140/epjc/s10052-018-6530-5}}.

\bibitem{Cepila:2016uku}
J.~Cepila, J.~G. Contreras, J.~D. Tapia~Takaki, {Energy dependence of
  dissociative $\mathrm{J/}\psi$ photoproduction as a signature of gluon
  saturation at the LHC}, Phys. Lett. B766 (2017) 186--191.
\newblock \href {http://arxiv.org/abs/1608.07559} {\path{arXiv:1608.07559}},
  \href {http://dx.doi.org/10.1016/j.physletb.2016.12.063}
  {\path{doi:10.1016/j.physletb.2016.12.063}}.

\bibitem{Cepila:2018zky}
J.~Cepila, J.~G. Contreras, M.~Krelina, J.~D. Tapia~Takaki, {Mass dependence of
  vector meson photoproduction off protons and nuclei within the
  energy-dependent hot-spot model}, Nucl. Phys. B934 (2018) 330--340.
\newblock \href {http://arxiv.org/abs/1804.05508} {\path{arXiv:1804.05508}},
  \href {http://dx.doi.org/10.1016/j.nuclphysb.2018.07.010}
  {\path{doi:10.1016/j.nuclphysb.2018.07.010}}.

\bibitem{Cepila:2017nef}
J.~Cepila, J.~G. Contreras, M.~Krelina, {Coherent and incoherent
  $\mathrm{J/}\psi$ photonuclear production in an energy-dependent hot-spot
  model}, Phys. Rev. C97~(2) (2018) 024901.
\newblock \href {http://arxiv.org/abs/1711.01855} {\path{arXiv:1711.01855}},
  \href {http://dx.doi.org/10.1103/PhysRevC.97.024901}
  {\path{doi:10.1103/PhysRevC.97.024901}}.

\bibitem{ATLAS:2017kwa}
T.~A. collaboration, {Photo-nuclear dijet production in ultra-peripheral Pb+Pb
  collisions}.

\bibitem{Aaboud:2017bwk}
M.~Aaboud, et~al., {Evidence for light-by-light scattering in heavy-ion
  collisions with the ATLAS detector at the LHC}, Nature Phys. 13~(9) (2017)
  852--858.
\newblock \href {http://arxiv.org/abs/1702.01625} {\path{arXiv:1702.01625}},
  \href {http://dx.doi.org/10.1038/nphys4208} {\path{doi:10.1038/nphys4208}}.

\bibitem{Aad:2019ock}
G.~Aad, et~al., {Observation of light-by-light scattering in ultraperipheral
  Pb+Pb collisions with the ATLAS detector}, Phys. Rev. Lett. 123~(5) (2019)
  052001.
\newblock \href {http://arxiv.org/abs/1904.03536} {\path{arXiv:1904.03536}},
  \href {http://dx.doi.org/10.1103/PhysRevLett.123.052001}
  {\path{doi:10.1103/PhysRevLett.123.052001}}.

\bibitem{Sirunyan:2018fhl}
A.~M. Sirunyan, et~al., {Evidence for light-by-light scattering and searches
  for axion-like particles in ultraperipheral PbPb collisions at
  $\sqrt{s_\mathrm{NN}} =$ 5.02 TeV}\href {http://arxiv.org/abs/1810.04602}
  {\path{arXiv:1810.04602}}, \href
  {http://dx.doi.org/10.1016/j.physletb.2019.134826}
  {\path{doi:10.1016/j.physletb.2019.134826}}.

\end{thebibliography}

\end{document}